\pdfoutput=1

\documentclass[12pt,a4paper]{article}

\usepackage{ifthen} 
\usepackage[normalem]{ulem} 
\newboolean{pdflatex}
\setboolean{pdflatex}{true} 

\newboolean{articletitles}
\setboolean{articletitles}{true} 

\newboolean{uprightparticles}
\setboolean{uprightparticles}{false} 

\def\paperauthors{LHCb collaboration} 
\def\paperasciititle{Measurement of CP violation in BdToDstD decays} 
\def\papertitle{Measurement of $CP$ violation in $B^0\to \D^{*\pm}\D^{\mp}$ decays} 
\def\paperkeywords{{High Energy Physics}, {LHCb}} 
\def\papercopyright{\the\year\ CERN for the benefit of the LHCb collaboration} 
\def\paperlicence{CC-BY-4.0 licence}
\def\paperlicenceurl{https://creativecommons.org/licenses/by/4.0/}

\usepackage[top=1in, bottom=1.25in, left=1in, right=1in]{geometry}

\usepackage{microtype}
\usepackage{lineno}  
\usepackage{xspace} 
\usepackage{caption} 

\usepackage{graphicx}  
\usepackage{color}
\usepackage{colortbl}
\graphicspath{{./}} 
\DeclareGraphicsExtensions{.pdf,.PDF,png,.PNG}

\usepackage{amsmath} 
\usepackage{amssymb}
\usepackage{amsfonts}
\usepackage{upgreek} 

\newcommand*\patchAmsMathEnvironmentForLineno[1]{%
\expandafter\let\csname old#1\expandafter\endcsname\csname #1\endcsname
\expandafter\let\csname oldend#1\expandafter\endcsname\csname
end#1\endcsname
 \renewenvironment{#1}%
   {\linenomath\csname old#1\endcsname}%
   {\csname oldend#1\endcsname\endlinenomath}%
}
\newcommand*\patchBothAmsMathEnvironmentsForLineno[1]{%
  \patchAmsMathEnvironmentForLineno{#1}%
  \patchAmsMathEnvironmentForLineno{#1*}%
}
\AtBeginDocument{%
\patchBothAmsMathEnvironmentsForLineno{equation}%
\patchBothAmsMathEnvironmentsForLineno{align}%
\patchBothAmsMathEnvironmentsForLineno{flalign}%
\patchBothAmsMathEnvironmentsForLineno{alignat}%
\patchBothAmsMathEnvironmentsForLineno{gather}%
\patchBothAmsMathEnvironmentsForLineno{multline}%
\patchBothAmsMathEnvironmentsForLineno{eqnarray}%
}

\usepackage{hyperxmp}

\usepackage[pdftex,
            pdfauthor={\paperauthors},
            pdftitle={\paperasciititle},
            pdfkeywords={\paperkeywords},
            pdfcopyright={Copyright (C) \papercopyright},
            pdflicenseurl={\paperlicenceurl}]{hyperref}

\usepackage[colorinlistoftodos,textsize=scriptsize]{todonotes}

\usepackage[all]{hypcap} 

\usepackage{xspace} 
\usepackage{upgreek}

\def\lhcb   {\mbox{LHCb}\xspace}

\def\babar  {\mbox{BaBar}\xspace}
\def\belle  {\mbox{Belle}\xspace}

\def\MagUp {\mbox{\em Mag\kern -0.05em Up}\xspace}

\ifthenelse{\boolean{uprightparticles}}%
{

 \def\Ppi         {\ensuremath{\uppi}\xspace}

 \def\Pphi        {\ensuremath{\upphi}\xspace}

 \def\Ppsi        {\ensuremath{\uppsi}\xspace}

 \def\PDelta      {\ensuremath{\Delta}\xspace}                 
 \def\PXi         {\ensuremath{\Xi}\xspace}                 
 \def\PLambda     {\ensuremath{\Lambda}\xspace}                 
 \def\PSigma      {\ensuremath{\Sigma}\xspace}                 
 \def\POmega      {\ensuremath{\Omega}\xspace}                 
 \def\PUpsilon    {\ensuremath{\Upsilon}\xspace}

 \def\PB      {\ensuremath{\mathrm{B}}\xspace}                 
                  
 \def\PD      {\ensuremath{\mathrm{D}}\xspace}

 \def\PJ      {\ensuremath{\mathrm{J}}\xspace}                 
 \def\PK      {\ensuremath{\mathrm{K}}\xspace}

 \def\PW      {\ensuremath{\mathrm{W}}\xspace}

 \def\Pb      {\ensuremath{\mathrm{b}}\xspace}                 
 \def\Pc      {\ensuremath{\mathrm{c}}\xspace}                 
 \def\Pd      {\ensuremath{\mathrm{d}}\xspace}

 \def\Pi      {\ensuremath{\mathrm{i}}\xspace}

 \def\Pp      {\ensuremath{\mathrm{p}}\xspace}

 \def\Ps      {\ensuremath{\mathrm{s}}\xspace}

 \def\thebaroffset{0.0em}
}
{

 \def\Ppi         {\ensuremath{\pi}\xspace}

 \def\Pphi        {\ensuremath{\phi}\xspace}

 \def\Ppsi        {\ensuremath{\psi}\xspace}                 
                  
 \mathchardef\PDelta="7101
 \mathchardef\PXi="7104
 \mathchardef\PLambda="7103
 \mathchardef\PSigma="7106
 \mathchardef\POmega="710A
 \mathchardef\PUpsilon="7107
                  
 \def\PB      {\ensuremath{B}\xspace}                 
                  
 \def\PD      {\ensuremath{D}\xspace}

 \def\PJ      {\ensuremath{J}\xspace}                 
 \def\PK      {\ensuremath{K}\xspace}

 \def\PW      {\ensuremath{W}\xspace}

 \def\Pb      {\ensuremath{b}\xspace}                 
 \def\Pc      {\ensuremath{c}\xspace}                 
 \def\Pd      {\ensuremath{d}\xspace}

 \def\Pi      {\ensuremath{i}\xspace}

 \def\Pp      {\ensuremath{p}\xspace}

 \def\Ps      {\ensuremath{s}\xspace}

 \def\thebaroffset{0.18em}
}
\newcommand{\offsetoverline}[2][\thebaroffset]{\kern #1\overline{\kern -#1 #2}}%

\makeatletter
\ifcase \@ptsize \relax
  \newcommand{\miniscule}{\@setfontsize\miniscule{4}{5}}
\or
  \newcommand{\miniscule}{\@setfontsize\miniscule{5}{6}}
\or
  \newcommand{\miniscule}{\@setfontsize\miniscule{5}{6}}
\fi
\makeatother

\DeclareRobustCommand{\optbar}[1]{\shortstack{{\miniscule (\rule[.5ex]{1.25em}{.18mm})}
  \\ [-.7ex] $#1$}}

\def\W      {{\ensuremath{\PW}}\xspace}

\def\dquark    {{\ensuremath{\Pd}}\xspace}

\def\squark    {{\ensuremath{\Ps}}\xspace}

\def\cquark    {{\ensuremath{\Pc}}\xspace}
\def\cquarkbar {{\ensuremath{\overline \cquark}}\xspace}

\def\bquark    {{\ensuremath{\Pb}}\xspace}
\def\bquarkbar {{\ensuremath{\overline \bquark}}\xspace}

\def\pion   {{\ensuremath{\Ppi}}\xspace}

\def\pip    {{\ensuremath{\pion^+}}\xspace}
\def\pim    {{\ensuremath{\pion^-}}\xspace}

\def\kaon    {{\ensuremath{\PK}}\xspace}

\def\KorKbar {\kern \thebaroffset\optbar{\kern -\thebaroffset \PK}{}\xspace}

\def\Kp      {{\ensuremath{\kaon^+}}\xspace}
\def\Km      {{\ensuremath{\kaon^-}}\xspace}

\def\KS      {{\ensuremath{\kaon^0_{\mathrm{S}}}}\xspace}

\newcommand{\phiz}{\ensuremath{\Pphi}\xspace}

\def\Dbar    {{\ensuremath{\offsetoverline{\PD}}}\xspace}
\def\D       {{\ensuremath{\PD}}\xspace}

\def\DorDbar {\kern \thebaroffset\optbar{\kern -\thebaroffset \PD}\xspace}
\def\Dz      {{\ensuremath{\D^0}}\xspace}
\def\Dzb     {{\ensuremath{\Dbar{}^0}}\xspace}
\def\Dp      {{\ensuremath{\D^+}}\xspace}
\def\Dm      {{\ensuremath{\D^-}}\xspace}
\def\Dpm     {{\ensuremath{\D^\pm}}\xspace}
\def\Dmp     {{\ensuremath{\D^\mp}}\xspace}
\def\Dstar   {{\ensuremath{\D^*}}\xspace}

\def\Dstarp  {{\ensuremath{\D^{*+}}}\xspace}
\def\Dstarm  {{\ensuremath{\D^{*-}}}\xspace}
\def\Dstarpm {{\ensuremath{\D^{*\pm}}}\xspace}

\def\Ds      {{\ensuremath{\D^+_\squark}}\xspace}
\def\Dsp     {{\ensuremath{\D^+_\squark}}\xspace}
\def\Dsm     {{\ensuremath{\D^-_\squark}}\xspace}

\def\B       {{\ensuremath{\PB}}\xspace}
\def\Bbar    {{\ensuremath{\offsetoverline{\PB}}}\xspace}

\def\BorBbar {\kern \thebaroffset\optbar{\kern -\thebaroffset \PB}\xspace}
\def\Bz      {{\ensuremath{\B^0}}\xspace}
\def\Bzb     {{\ensuremath{\Bbar{}^0}}\xspace}
\def\Bd      {{\ensuremath{\B^0}}\xspace}
\def\Bdb     {{\ensuremath{\Bbar{}^0}}\xspace}
\def\BdorBdbar {\kern \thebaroffset\optbar{\kern -\thebaroffset \Bd}\xspace}
\def\Bu      {{\ensuremath{\B^+}}\xspace}

\def\Bs      {{\ensuremath{\B^0_\squark}}\xspace}

\def\BsorBsbar {\kern \thebaroffset\optbar{\kern -\thebaroffset \Bs}\xspace}

\def\Bds     {{\ensuremath{\B_{(\squark)}^0}}\xspace}

\def\jpsi     {{\ensuremath{{\PJ\mskip -3mu/\mskip -2mu\Ppsi\mskip 2mu}}}\xspace}

\def\Y#1S{\ensuremath{\PUpsilon{(#1S)}}\xspace}

\def\proton      {{\ensuremath{\Pp}}\xspace}

\def\Lz          {{\ensuremath{\PLambda}}\xspace}

\def\LorLbar     {\kern \thebaroffset\optbar{\kern -\thebaroffset \PLambda}\xspace}

\def\Lc          {{\ensuremath{\Lz^+_\cquark}}\xspace}

\def\Lb           {{\ensuremath{\Lz^0_\bquark}}\xspace}

\newcommand{\decay}[2]{\ensuremath{#1\!\to #2}\xspace} 

\def\to                 {\ensuremath{\rightarrow}\xspace}

\def\CP                {{\ensuremath{C\!P}}\xspace}

\newcommand{\dmd}{{\ensuremath{\Delta m_{\dquark}}}\xspace}

\newcommand{\DGd}{{\ensuremath{\Delta\Gamma_{\dquark}}}\xspace}

\newcommand{\mistag}{\ensuremath{\omega}\xspace}

\newcommand{\etag}{{\ensuremath{\varepsilon_{\mathrm{tag}}}}\xspace}

\newcommand{\effD}{{\ensuremath{\etag D^2}}\xspace}

\def\AT#1     {\ensuremath{A_{\mathrm{T}}^{#1}}\xspace}           

\def\C#1      {\ensuremath{\mathcal{C}_{#1}}\xspace}                       
\def\Cp#1     {\ensuremath{\mathcal{C}_{#1}^{'}}\xspace}                    
\def\Ceff#1   {\ensuremath{\mathcal{C}_{#1}^{\mathrm{(eff)}}}\xspace}        
\def\Cpeff#1  {\ensuremath{\mathcal{C}_{#1}^{'\mathrm{(eff)}}}\xspace}       
\def\Ope#1    {\ensuremath{\mathcal{O}_{#1}}\xspace}                       
\def\Opep#1   {\ensuremath{\mathcal{O}_{#1}^{'}}\xspace}                    

\newcommand{\ket}[1]{\ensuremath{|#1\rangle}}              

\newcommand{\nospaceunit}[1]{\ensuremath{\text{#1}}}       
\newcommand{\aunit}[1]{\ensuremath{\text{\,#1}}}       

\newcommand{\tev}{\aunit{Te\kern -0.1em V}\xspace}
\newcommand{\gev}{\aunit{Ge\kern -0.1em V}\xspace}
\newcommand{\mev}{\aunit{Me\kern -0.1em V}\xspace}
\newcommand{\kev}{\aunit{ke\kern -0.1em V}\xspace}
\newcommand{\ev}{\aunit{e\kern -0.1em V}\xspace}
\newcommand{\mevc}{\ensuremath{\aunit{Me\kern -0.1em V\!/}c}\xspace}
\newcommand{\gevc}{\ensuremath{\aunit{Ge\kern -0.1em V\!/}c}\xspace}
\newcommand{\mevcc}{\ensuremath{\aunit{Me\kern -0.1em V\!/}c^2}\xspace}
\newcommand{\gevcc}{\ensuremath{\aunit{Ge\kern -0.1em V\!/}c^2}\xspace}

\def\mm   {\aunit{mm}\xspace}

\def\mum  {\ensuremath{\,\upmu\nospaceunit{m}}\xspace}

\def\fb   {\ensuremath{\aunit{fb}}\xspace}
\def\invfb   {\ensuremath{\fb^{-1}}\xspace}

\def\ps   {\ensuremath{\aunit{ps}}\xspace}
\def\fs   {\aunit{fs}}

\newcommand{\chisq}{\ensuremath{\chi^2}\xspace}

\newcommand{\chisqip}{\ensuremath{\chi^2_{\text{IP}}}\xspace}

\def\gsim{{~\raise.15em\hbox{$>$}\kern-.85em
          \lower.35em\hbox{$\sim$}~}\xspace}
\def\lsim{{~\raise.15em\hbox{$<$}\kern-.85em
          \lower.35em\hbox{$\sim$}~}\xspace}

\def\PDF {PDF\xspace}

\def\sPlot{\mbox{\em sPlot}\xspace}

\def\pt         {\ensuremath{p_{\mathrm{T}}}\xspace}

\def\ptot       {\ensuremath{p}\xspace}

\def\evtgen     {\mbox{\textsc{EvtGen}}\xspace}

\def\geant      {\mbox{\textsc{Geant4}}\xspace}

\def\photos     {\mbox{\textsc{Photos}}\xspace}

\def\pythia     {\mbox{\textsc{Pythia}}\xspace}

\def\tell1  {TELL1\xspace}
\def\ukl1   {UKL1\xspace}

\newcommand{\ie}{\mbox{\itshape i.e.}\xspace}

\def\DstD    {\ensuremath{\Dstarpm\Dmp}\xspace}
\newcommand{\BdToDstD}{\mbox{\decay{\Bd}{\Dstarpm\Dmp}}\xspace}
\newcommand{\BsToDstD}{\mbox{\decay{\Bs}{\Dstarpm\Dmp}}\xspace}

\newcommand{\BsToDstDst}{\mbox{\decay{\Bs}{\Dstarp\Dstarm}}\xspace}
\newcommand{\BdToDstDst}{\mbox{\decay{\Bd}{\Dstarp\Dstarm}}\xspace}

\newcommand{\BdToDsD}{\mbox{\decay{\Bd}{\Dsp\Dm}}\xspace}
\newcommand{\BsToDsD}{\mbox{\decay{\Bs}{\Dsm\Dp}}\xspace}

\newcommand{\BdToDstDs}{\mbox{\decay{\Bd}{\Dsp\Dstarm}}\xspace}

\newcommand{\BsToDstDs}{\mbox{\decay{\Bs}{\Dsm\Dstarp}}\xspace}

\newcommand{\DzToKpi}{\mbox{\decay{\Dz}{\Km\pip}}\xspace}
\newcommand{\DzToKpipipi}{\mbox{\decay{\Dz}{\Km\pim\pip\pip}}\xspace}
\newcommand{\DToKpipi}{\mbox{\decay{\Dm}{\Kp\pim\pim}}\xspace}
\newcommand{\DpToKpipi}{\mbox{\decay{\Dp}{\Km\pip\pip}}\xspace}
\newcommand{\DsToKKpi}{\mbox{\decay{\Ds}{\Kp\Km\pip}}\xspace}

\newcommand{\bToccs}{\mbox{\decay{\bquark}{\cquark\cquarkbar\squark}}\xspace}
\newcommand{\bToccd}{\mbox{\decay{\bquark}{\cquark\cquarkbar\dquark}}\xspace}

\def\Araw       {\ensuremath{{\cal A}_{\rm raw}}\xspace} 
\def\Adet		{\ensuremath{{\cal A}_{\rm det}}\xspace}
\def\ArawKpiU   {\ensuremath{{\cal A}_{\rm raw}^{K\pi,\text{Run1}}}\xspace}
\def\ArawKpipipiU {\ensuremath{{\cal A}_{\rm raw}^{K\pi\pi\pi,\text{Run1}}}\xspace}
\def\ArawKpiD     {\ensuremath{{\cal A}_{\rm raw}^{K\pi,\text{Run2}}}\xspace}
\def\ArawKpipipiD {\ensuremath{{\cal A}_{\rm  raw}^{K\pi\pi\pi,\text{Run2}}}\xspace}

\def\ADstD		{\ensuremath{{\cal A}_{\Dstar\D}}\xspace}
\def\SDstD		{\ensuremath{S_{\Dstar\D}}\xspace}
\def\DeltaSDstD	{\ensuremath{\Delta S_{\Dstar\D}}\xspace}
\def\CDstD		{\ensuremath{C_{\Dstar\D}}\xspace}
\def\DeltaCDstD	{\ensuremath{\Delta C_{\Dstar\D}}\xspace}
\def\Kpi			{\ensuremath{\Km\pip}\xspace}
\def\Kpipipi		{\ensuremath{\Km\pim\pip\pip}\xspace}
\usepackage{cite} 
\usepackage{mciteplus}

\usepackage{longtable} 
\usepackage{rotating} 
\usepackage{multirow} 
\begin{document}

\renewcommand{\thefootnote}{\fnsymbol{footnote}}
\setcounter{footnote}{1}


\begin{titlepage}
\pagenumbering{roman}

\vspace*{-1.5cm}
\centerline{\large EUROPEAN ORGANIZATION FOR NUCLEAR RESEARCH (CERN)}
\vspace*{1.5cm}
\noindent
\begin{tabular*}{\linewidth}{lc@{\extracolsep{\fill}}r@{\extracolsep{0pt}}}
\ifthenelse{\boolean{pdflatex}}
{\vspace*{-1.5cm}\mbox{\!\!\!\includegraphics[width=.14\textwidth]{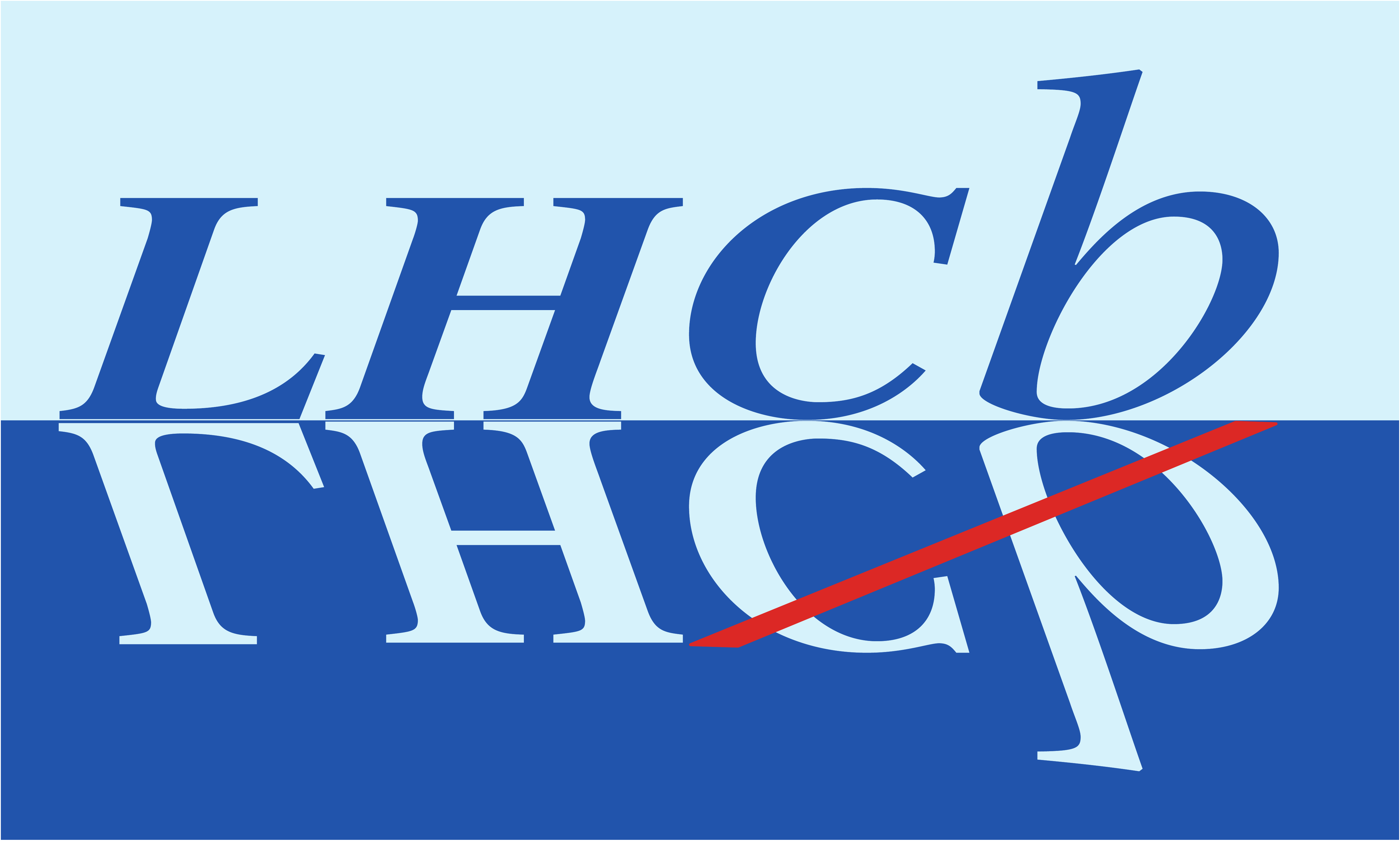}} & &}%
{\vspace*{-1.2cm}\mbox{\!\!\!\includegraphics[width=.12\textwidth]{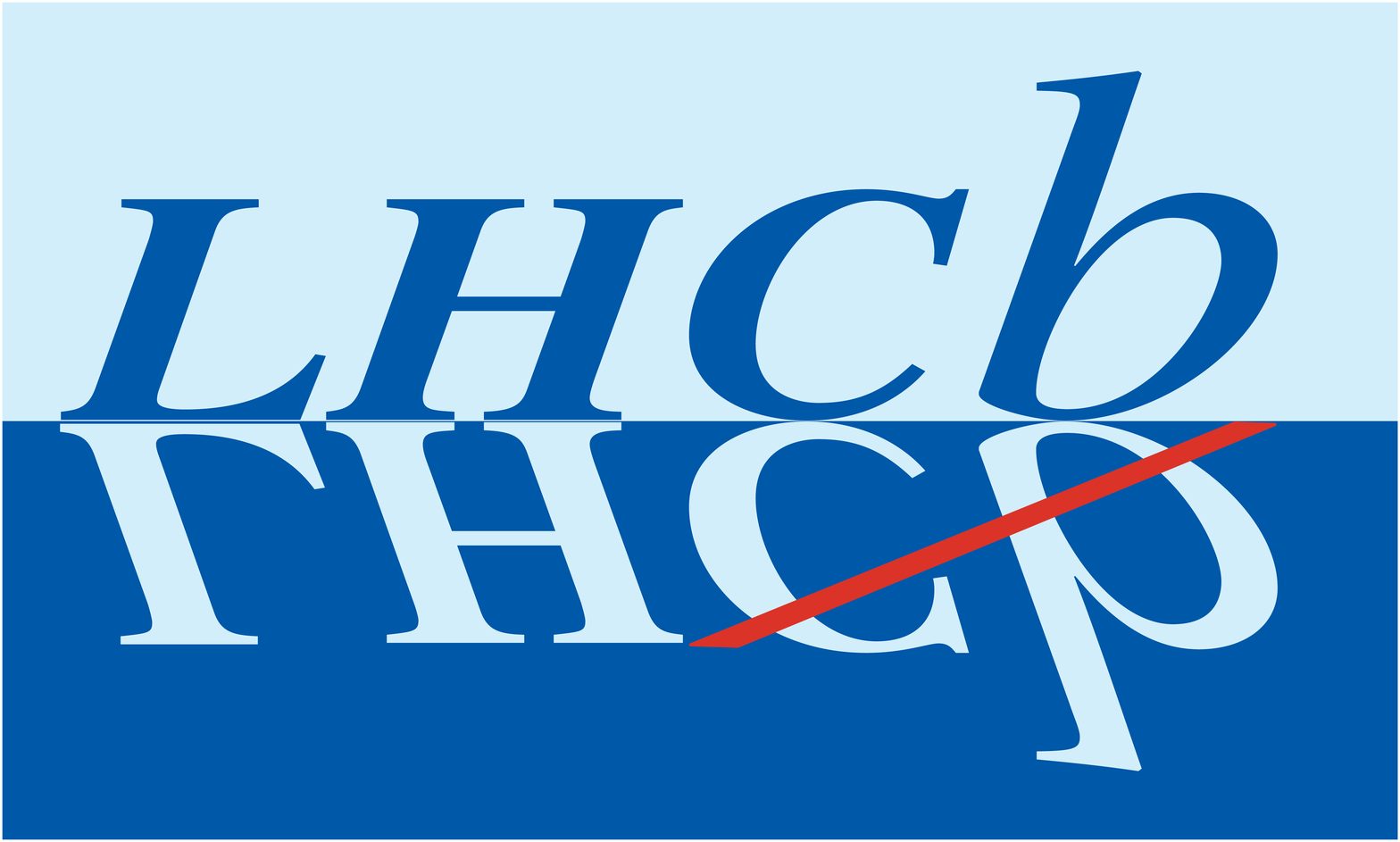}} & &}%
\\
 & & CERN-EP-2019-264 \\  
 & & LHCb-PAPER-2019-036 \\  
 & & 3 April 2020 \\ 
 & & \\
\end{tabular*}

\vspace*{4.0cm}

{\normalfont\bfseries\boldmath\huge
\begin{center}
  \papertitle 
\end{center}
}

\vspace*{2.0cm}

\begin{center}
\paperauthors\footnote{Authors are listed at the end of this paper.}
\end{center}

\vspace{\fill}

\begin{abstract}
  \noindent
    The decay-time-dependent \ensuremath{C\!P} asymmetry in $B^0 \to D^{*\pm}D^{\mp}$ decays is measured using a data set corresponding to an integrated luminosity of $9$\ensuremath{\fb^{-1}} recorded by the \lhcb detector in proton-proton collisions at centre-of-mass energies of 7, 8 and $13$\aunit{Te\kern -0.1em V}. 
    The \ensuremath{C\!P} parameters are measured as
    \begin{equation*}
      \begin{split}
        S_{\D^*\D} &=                              {-0.861} \pm {0.077} \,\text{(stat)} \pm {0.019} \,\text{(syst)}\,, \\
         \Delta S_{\D^*\D} &=       \phantom{+}     {0.019} \pm {0.075} \,\text{(stat)} \pm {0.012} \,\text{(syst)} \,,\\ 
        C_{\D^*\D} &=                              {-0.059} \pm {0.092} \,\text{(stat)} \pm {0.020} \,\text{(syst)} \,,\\
        \Delta C_{\D^*\D} &=                        {-0.031} \pm {0.092} \,\text{(stat)} \pm {0.016} \,\text{(syst)} \,,\\
        {\cal A}_{\D^*\D} &=         \phantom{+}          {0.008} \pm {0.014} \,\text{(stat)} \pm  0.006 \,\text{(syst)}\,.
      \end{split}
    \end{equation*}
    The analysis provides the most precise single measurement of \ensuremath{C\!P} violation in this decay channel to date. All parameters are consistent with their current world average values.   
\end{abstract}

\vspace*{2.0cm}

\begin{center}
  Published in JHEP 03 (2020)  147. 
\end{center}

\vspace{\fill}

{\footnotesize 
\centerline{\copyright~\papercopyright. \href{\paperlicenceurl}{\paperlicence}.}}
\vspace*{2mm}

\end{titlepage}


\newpage
\setcounter{page}{2}
\mbox{~}
%
%
%
%

\cleardoublepage


\renewcommand{\thefootnote}{\arabic{footnote}}
\setcounter{footnote}{0}



\pagestyle{plain} 
\setcounter{page}{1}
\pagenumbering{arabic}


%

\section{Introduction}
\label{sec:Introduction}
 In the hadronic sector of the Standard Model (SM), \CP violation originates from an irreducible complex phase in the Cabibbo--Kobayashi--Maskawa matrix that describes the mixing of the quark mass eigenstates into weak-interaction eigenstates~\cite{Cabibbo:1963yz,Kobayashi:1973fv}.
Interference caused by a weak-phase difference between the \Bz--\Bzb oscillation and the decay amplitudes leads to a \CP asymmetry in the decay-time distributions of  \Bz and \Bzb mesons. 
Decays involving \bToccs tree transitions at leading order, such as \decay{\Bd}{\jpsi\KS},\footnote{Charge-conjugated processes are implicitly included in the following, unless specified.} are sensitive to the weak phase $2\beta$, where  $\beta\equiv\arg[-(V_{cd}^{}V_{cb}^{\ast})/(V_{td}^{}V_{tb}^{\ast})]$ is one of the angles of the Unitarity Triangle.
Measurements of this phase were performed by several experiments using different channels ~\cite{HFLAV16}.
The same phase appears in \bToccd transitions, which contribute to \BdToDstD decays, when the leading-order colour-favoured tree diagram is considered.
However, \BdToDstD decays can also proceed through several other decay diagrams, that include penguin, \W-exchange and annihilation topologies, where additional contributions to \CP violation both from the SM and new physics (NP) may arise. 
Tests of the SM have been performed by relating \CP asymmetries and branching fractions of different decay modes of neutral and charged beauty mesons to two charm mesons \cite{PhysRevD.91.034027,Bel2015}.

Each of the \Dstarp\Dm and \Dstarm\Dp final states are accessible from both \Bz and \Bzb mesons. The time-dependent decay rates for the four configurations of initial \B flavour and final states  can be written as
\begin{equation}
\label{eq:cpvdstd1}
\begin{split}
\frac{\mathrm{d}\Gamma_{\Bzb,f}(t)}{\mathrm{d}t} &= \frac{{e}^{-t/\tau_d}}{8\tau_d}(1+{\cal A}_{f\bar{f}})\Bigl[ 1+ S_f \sin(\Delta m_d t)-C_f \cos(\Delta m_d t)\Bigr]  , \\
\frac{\mathrm{d}\Gamma_{\Bz,f}(t)}{\mathrm{d}t} &= \frac{{e}^{-t/\tau_d}}{8\tau_d}(1+{\cal A}_{f\bar{f}})\Bigl[ 1- S_f \sin(\Delta m_d t)+C_f \cos(\Delta m_d t)\Bigr]  , \\
\frac{\mathrm{d}\Gamma_{\Bzb,\bar{f}}(t)}{\mathrm{d}t} &= \frac{{e}^{-t/\tau_d}}{8\tau_d}(1-{\cal A}_{f\bar{f}})\Bigl[ 1+ S_{\bar{f}} \sin(\Delta m_d t)-C_{\bar{f}} \cos(\Delta m_d t)\Bigr]  , \\
\frac{\mathrm{d}\Gamma_{\Bz,\bar{f}}(t)}{\mathrm{d}t} &= \frac{{e}^{-t/\tau_d}}{8\tau_d}(1-{\cal A}_{f\bar{f}})\Bigl[ 1- S_{\bar{f}} \sin(\Delta m_d t) + C_{\bar{f}} \cos(\Delta m_d t)\Bigr] ,
\end{split}
\end{equation}
where  $f=\Dstarp\Dm$ and $\bar{f}=\Dstarm\Dp$.
The parameter ${\cal A}_{f\bar{f}}$ represents the overall asymmetry in the production of the $f$ and $\bar{f}$ final states and is defined as
\begin{equation}
\label{eq:cpvdstdasymm}
{\cal A}_{f\bar{f}} = \frac{\Bigl( \lvert A_f \rvert ^2 + \lvert \bar{A}_f\rvert^2\Bigr)-\Bigl( \lvert A_{\bar{f}}\rvert^2+\lvert \bar{A}_{\bar{f}}\rvert^2\Bigr)}{\Bigl( \lvert A_f\rvert^2 + \lvert \bar{A}_f\rvert^2\Bigr)+\Bigl( \lvert A_{\bar{f}}\rvert^2+\lvert \bar{A}_{\bar{f}}\rvert^2\Bigr)},
\end{equation}
with  $A_f$ ($A_{\bar{f}}$) and ${\bar{A}}_f$ (${\bar{A}}_{\bar{f}}$) indicating  the amplitudes of the decay of a  \Bz and a \Bzb meson to final state $f$ ($\bar{f}$). 
Here, $\tau_d$ is the \Bz lifetime and \dmd is the mass difference of the two \Bd mass eigenstates, 
which are assumed to have the same decay width~\cite{HFLAV16}. 
Introducing $q$ and $p$ to describe the relation between the mass and flavour eigenstates, $\ket{B_{\rm H,L}}=p \ket{\Bz} \pm q \ket{\Bzb}$, the parameters $S_f$ and $C_f$ are defined as
\begin{equation}
	\label{eq:ckm12}
	S_f=\frac{2\mathcal{I}m\lambda_f}{1+\lvert \lambda_f \rvert^2} ,\hspace{1cm} C_f=\frac{1-\lvert\lambda_f\rvert^2}{1+\lvert\lambda_f \rvert^2} ,\hspace{1cm}\lambda_f=\frac{q}{p}\frac{\bar{A}_f}{A_f}\, ,
	\end{equation}
with analogous definitions holding for $S_{\bar{f}}$, and $C_{\bar{f}}$. 
By combining these parameters, the \CP observables for \BdToDstD decays can be defined as~\cite{HFLAV16} 
\begin{equation}
\label{eq:cpvdstd4}
\begin{split}
\SDstD &= \frac{1}{2} ( S_f+ S_{\bar f}) , \\
\DeltaSDstD &= \frac{1}{2} ( S_f - S_{\bar f}) , \\
\CDstD &= \frac{1}{2} (C_f+ C_{\bar f}) , \\
\DeltaCDstD &= \frac{1}{2} ( C_f - C_{\bar f}), \\
\ADstD &= {\cal A}_{f\bar{f}}.
\end{split}
\end{equation}
In absence of \CP violation, \SDstD and \CDstD vanish. While \DeltaSDstD is related to the relative strong phase between the decay amplitudes, the parameter \DeltaCDstD is a measure of how flavour specific the decay mode is.
For a flavour-specific decay only one final state is accessible for each flavour of the decaying neutral \B meson, $\DeltaCDstD=\pm1$ and no \CP violation in the interference between decays with and without mixing is possible. Decays with $\DeltaCDstD=0$ present the highest sensitivity to mixing-induced \CP violation. 
In the case of \BdToDstD decays, if the contribution
of higher-order SM processes and NP
are negligible, the amplitudes for $\Bz \to \Dstarp\Dm$ and $\Bz \to \Dstarm\Dp$ have the same hadronic phase and magnitude. As a result, \ADstD, \CDstD, \DeltaCDstD and \DeltaSDstD vanish and $\SDstD = \sin(2\beta)$.
Theoretical models, based on QCD factorization and heavy quark symmetry, estimate the contribution of penguin amplitudes in \BdToDstD to be up to a few percent~\cite{PhysLettB443.365,PhysRevD61.014010}. 

By combining Eqs.~\ref{eq:cpvdstd1} and~\ref{eq:cpvdstd4} the  decay rate can be rewritten as
\begin{alignat}{14}
\label{eq:cpvdstd5}
\frac{\mathrm{d}\Gamma(t)}{\mathrm{d}t} =& \frac{{e}^{-t/\tau_d}}{8\tau_d}(1+ r\ADstD )\times \\ \notag
&\Bigl[ 1- d(\SDstD+r\DeltaSDstD)&\sin(\Delta m t) + d (\CDstD+ r\DeltaCDstD) \cos(\Delta m t)\Bigr]\,  ,
\end{alignat}
where $d$ takes values $+1$ ($-1$) for mesons whose initial flavour is \Bz (\Bzb) and $r$ takes values $+1$ ($-1$) for the final states $f$ ($\bar{f}$).

This paper reports the first measurement of \CP violation in \BdToDstD decays at the \lhcb experiment.
The measurement is based on a sample of $pp$ collision data corresponding to integrated luminosities of 1 and 2\invfb at centre-of-mass energies of 7 and 8\tev (referred to as Run 1) and of 6\invfb at 13\tev (Run 2), recorded by the \lhcb experiment between 2011 and 2018. Previous measurements with this \Bz decay mode have been performed by the \babar \cite{Aubert:2008ah}  and \belle experiments \cite{Rohrken:2012ta}.

In this analysis the  \BdToDstD candidates are reconstructed through the subsequent decays \decay{\Dm}{\Kp\pim\pim} and \decay{\Dstarp}{\Dz\pip}.
For the \Dz meson, the \decay{\Dz}{\Km\pip\pip\pim} and \decay{\Dz}{\Km\pip} decay modes are used.
The analysis proceeds as follows: \BdToDstD candidates, reconstructed in the two \Dz decay modes and the two data-taking periods, are selected and analysed separately, as outlined in  Sec.~\ref{sec:Selection}.
The signal contribution is determined in each of the four samples with fits to the \Bz mass distributions, as described in Sec.~\ref{sec:Massfit}.
A key ingredient for measurements of \CP violation in time-dependent analyses is the determination of the flavour of the neutral \B mesons by means of tagging algorithms, described in Sec.~\ref{sec:Tagging}. 
The evaluation of instrumental asymmetries that affect the measurement of the overall \CP charge asymmetry \ADstD is discussed in Sec.~\ref{sec:Asymmetries}.
A simultaneous fit to the \Bz decay-time distributions of the four samples 
is performed to determine the \CP parameters, as described in Sec.~\ref{sec:Decaytimefit}.
The estimate of the systematic uncertainties is presented in Sec.~\ref{sec:Systematics} and finally, conclusions are drawn in Sec.~\ref{sec:conclusion}.

\section{Detector and simulation}
\label{sec:Detector}
The \lhcb  detector~\cite{LHCb-DP-2008-001,LHCb-DP-2014-002} is a
single-arm forward spectrometer covering the pseudorapidity range $2-5$, designed to study particles containing \bquark\ or \cquark\ quarks. The detector includes a high-precision tracking system consisting of a silicon-strip vertex detector surrounding the $pp$ interaction region~\cite{LHCb-DP-2014-001},
a large-area silicon-strip detector located upstream of a dipole magnet with a bending power of about
$4{\mathrm{\,Tm}}$, and three stations of silicon-strip detectors and straw
drift tubes~\cite{LHCb-DP-2013-003,LHCb-DP-2017-001} placed downstream of the magnet.
The tracking system provides a measurement of the momentum, \ptot, of charged particles with
a relative uncertainty that varies from 0.5\% at low momentum to 1.0\% at 200\gevc.
During the data taking, the polarity of the magnetic field  was periodically
reversed to reduce the residual detection asymmetries that affect the determination of charge asymmetries.
The minimum distance of a track to a primary collision vertex (PV), the impact parameter (IP), 
is measured with a resolution of $(15+29/\pt)\mum$,
where \pt is the component of the momentum transverse to the beam, in\,\gevc.
Different types of charged hadrons are distinguished using information
from two ring-imaging Cherenkov detectors~\cite{LHCb-DP-2012-003}.
Photons, electrons and hadrons are identified by a calorimeter system consisting of
scintillating-pad and preshower detectors, an electromagnetic
and a hadronic calorimeter. Muons are identified by a
system composed of alternating layers of iron and multiwire
proportional chambers~\cite{LHCb-DP-2012-002}.

Simulated data samples are used to model the effects of the detector acceptance and the imposed selection requirements. Samples of signal decays are produced in order to determine inputs for the analysis, such as the parametrisation of the mass distribution and  the decay-time resolution model. {Multibody  \Dz and \Dp decays are
modelled in the simulation according to the previously measured resonant structures~\cite{LHCb-PAPER-2017-040,PDG2018}.}
Samples of the most relevant background from partially reconstructed and misidentified \B meson decays, as well as specific \B decays useful for studies related to flavour tagging, are also produced. 
In the simulation, $pp$ collisions are generated using \pythia~\cite{Sjostrand:2006za,*Sjostrand:2007gs}  with a specific \lhcb configuration~\cite{LHCb-PROC-2010-056}.  
Decays of unstable particles
are described by \evtgen~\cite{Lange:2001uf}, in which final-state
radiation is generated using \photos~\cite{Golonka:2005pn}. The
interaction of the generated particles with the detector, and its response, are implemented using the \geant
toolkit~\cite{Allison:2006ve, *Agostinelli:2002hh} as described in
Ref.~\cite{LHCb-PROC-2011-006}.
\section{Selection}
\label{sec:Selection}
The online event selection is performed by a trigger~\cite{LHCb-DP-2012-004}, which consists of a hardware stage, based on information from the calorimeter and muon systems, followed by a software stage, which applies a full event reconstruction.
At the hardware-trigger stage, events are required to have a muon with high \pt or a hadron, photon or electron with high transverse energy in the calorimeters. 
The software trigger requires a two-, three- or four-track
secondary vertex with a large sum of the transverse momenta of
the tracks and a significant displacement from the primary $pp$
interaction vertices. 
A multivariate algorithm~\cite{BBDT} is used for the identification of secondary vertices consistent with the decay
of a \bquark hadron.

In the offline selection, the \Dstarp, \Dz and \Dm candidates are reconstructed through their decays into the selected final-state particles whose tracks are required to have good quality, exceed threshold values on $p$ and \pt and satisfy loose particle identification (PID) criteria,  mostly relying on the Cherenkov detectors information. These tracks are also required to have a \chisqip\ value with respect to any PV greater than four, where \chisqip\ is defined as the difference in the vertex-fit \chisq of a given PV reconstructed with and without the particle being considered.
The distance of closest approach between all possible combinations of particles forming a common vertex should be smaller than $0.5 \mm$ and the vertex should be downstream of the PV.
The invariant mass of
\Dz (\Dm) meson candidates is required to lie within $\pm 40\mevcc$ ($\pm 50\mevcc$) of the known value~\cite{PDG2018}, while the difference of the \Dstarp and \Dz invariant masses is required to be smaller than $150\mevcc$. {These windows correspond to about $\pm 5$ times the mass resolutions.}
Candidate \Bd mesons are reconstructed from \Dstarp and \Dm meson candidates that form a common vertex. The scalar sum of transverse momenta of the all final-state particles should exceed 5000\mevc and the momentum direction of the \Bd meson should point to its associated PV. If more than one PV is reconstructed in an event, the associated PV is that with respect to which the signal \Bz candidate has the smallest $\chisqip$.

Background can be due to the misidentification of one hadron in the charged \D decay chain.
To suppress $\Lb \to \Lc \Dstarm$ with $\Lc \to \proton \Kp \pim$
and \BdToDstDs with \DsToKKpi decays, mass vetoes are applied.
The pion from the \DpToKpipi reconstructed decay with the higher \pt is assumed to be a proton (kaon) and the candidate is rejected if the recomputed invariant mass is within $\pm 25 \mevcc$ of the known \Lc (\Dsp) mass and the PID requirement for the alternative particle assignment is satisfied.
To further reduce background contributions of $\phi\to\Kp\Km$ from \Dsm decays, the kaon mass hypothesis is assigned to the pion with the higher \pt in the \DToKpipi reconstructed decay and the candidate \Dm is rejected if the invariant mass of the kaon pair is compatible with the $\phi$ mass within $\pm 10\mevcc$.
These vetoes reduce the \Lc and \phiz background contributions to a negligible level. 
The background due to \decay{\Bd}{\Dsp \Dstarm} decay  is only partially suppressed by the \Dsp veto which includes only loose selection criteria to retain high signal efficiency.

Single-charm \B decays such as \decay{\Bds}{\Dstarm h^-h^+h^+}, where the three hadrons are not produced in a \Dp decay, but directly originate from the \Bd decay, are another potential source of background. To reject \decay{\Bz}{\Dstarm \pim\pip\pip} decays with a pion misidentified as a kaon,  the $\Dstarm \pim\pip\pip$  invariant mass is calculated with the pion mass assigned to the kaon. The candidate is rejected  if the mass is within $\pm 40\mevcc$ of the known \Bz mass, and either the kaon candidate has a high probability to be a pion or the $\chi^2$ of the flight distance of the \Dp with respect to the \Bz decay vertex is less than four. Background arising from possible \decay{\Bs}{\Dstarm \Km\pip\pip} decays is suppressed by rejecting candidates with the ${\Dstarm\Km\pip\pip}$ mass within $\pm 25\mevcc$ of the known \Bs mass
and the \Dp decay vertex reconstructed upstream the \Bz decay vertex, 
or the \chisq of the flight distance of the \Dp candidate with respect to the \Bz decay vertex smaller than two. This veto is applied only to \Dz decaying into \Kpipipi, as no excess is observed for the \Kpi sample.
The invariant-mass distribution of signal candidates is not significantly modified by the vetoes.

In order to separate further the \BdToDstD signal candidates from the combinatorial background, a boosted decision tree~(BDT) utilising the AdaBoost method~\cite{Breiman,AdaBoost} implemented in the TMVA toolkit~\cite{Hocker:2007ht,*TMVA4} is used.
To train the BDT, simulated \BdToDstD candidates are used as a proxy for signal, whereas candidates in data with invariant mass in the upper sideband are used as proxy for background. The upper sideband is defined as $5400<m_{\Dstarpm\Dmp}<6000$\mevcc ($5600<m_{\Dstarpm\Dmp}<6000$\mevcc) for the \Dz final states \Kpi and \Kpipipi. {A two-folding procedure is applied to avoid overtraining for the \Kpi candidates in the mass range $5400<m_{\Dstarpm\Dmp}<5600$\mevcc, which are both used for the BDT training and the mass fit.
}
Separate classifiers are trained for each final state and data taking period.
Various kinematic and topological quantities are used in the BDT to exploit the features of the signal decay in order to distinguish it from background, namely the transverse momentum of the \Bz, \Dstarp, \Dm mesons and of the \Dm and \Dz  decay products;  the decay-time significance of the \Dm and \Dz mesons, and their flight distance \chisq with respect to the associated PV; the \Bz \chisqip; the angles formed by each of the \Dm decay products and the \Dm direction; the angle formed by the pion from the \Dz decay and the \Dz direction, 
and the angle formed by the \Dstarp and the \Bz mesons.
The PID probabilities of the kaon and pion in the final state are also used in the BDT. 
The requirement on the output of each BDT classifiers is chosen to minimise 
the uncertainties on the \CP parameters \SDstD and \CDstD.

A kinematic fit to the \Bd decay chain with constraints on the masses of all the charm mesons and on the PV is performed to {significantly }improve {($\sim 60\%$)} the resolution on the invariant mass of the \Bd candidate~\cite{Hulsbergen:2005pu}, while
the \Bz decay time is calculated using the same fit with only the PV constraint in order  to avoid correlations of the decay time with the invariant mass.
Candidate \Bz mesons are retained if their invariant mass and decay time are in the ranges $5000-5600$\mevcc and $0.3-10.3$\ps, respectively, where the lower boundary of the decay time is set to reduce promptly produced background. 
After the selection a fraction of events below 5\% have multiple candidates. 
In these cases a single candidate is randomly selected with negligible change in the final result. 
\section{Mass fit} \label{sec:Massfit}
\begin{figure}[t]
\centering
\includegraphics[width=0.49\textwidth]{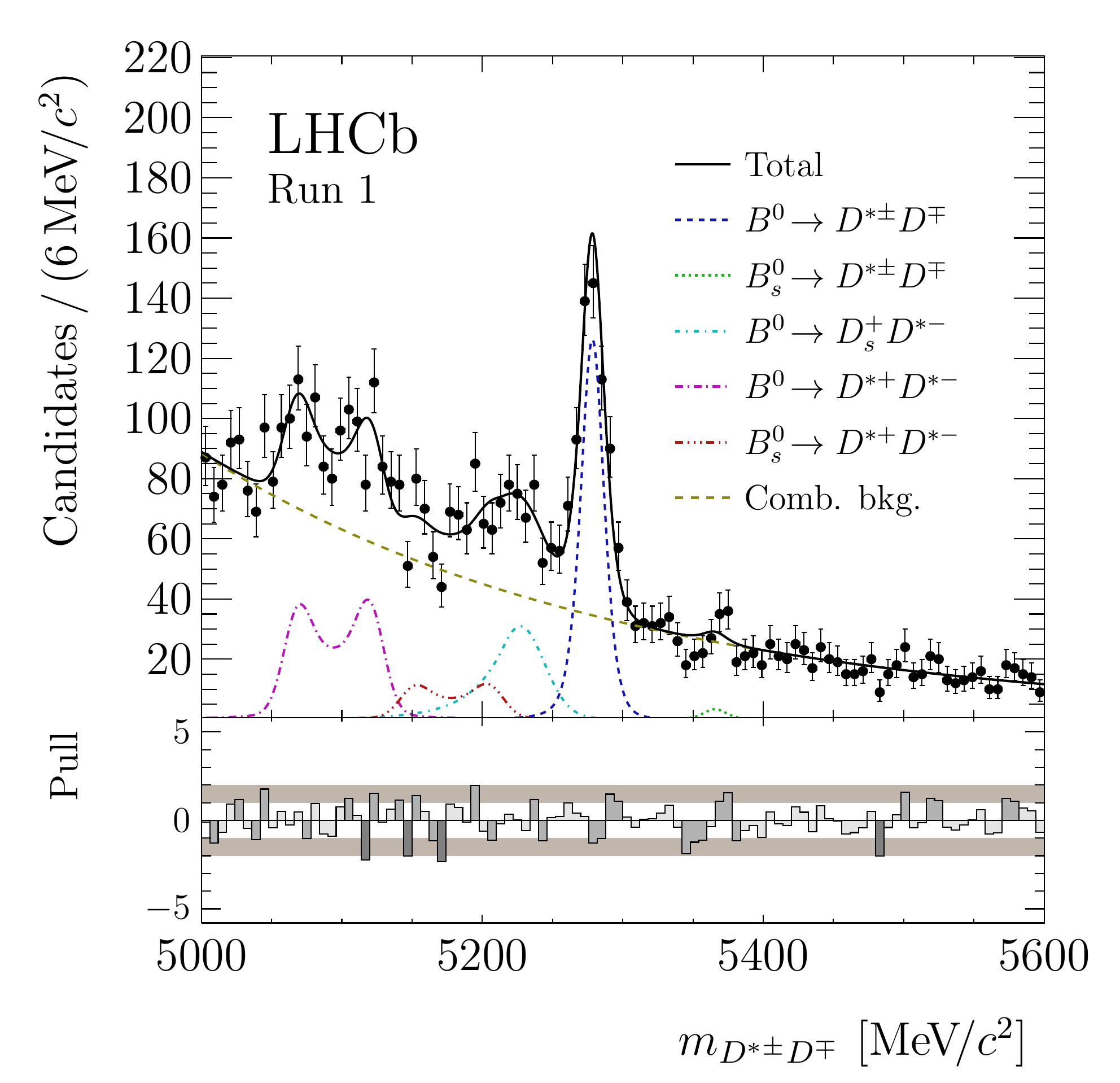}
\includegraphics[width=0.49\textwidth]{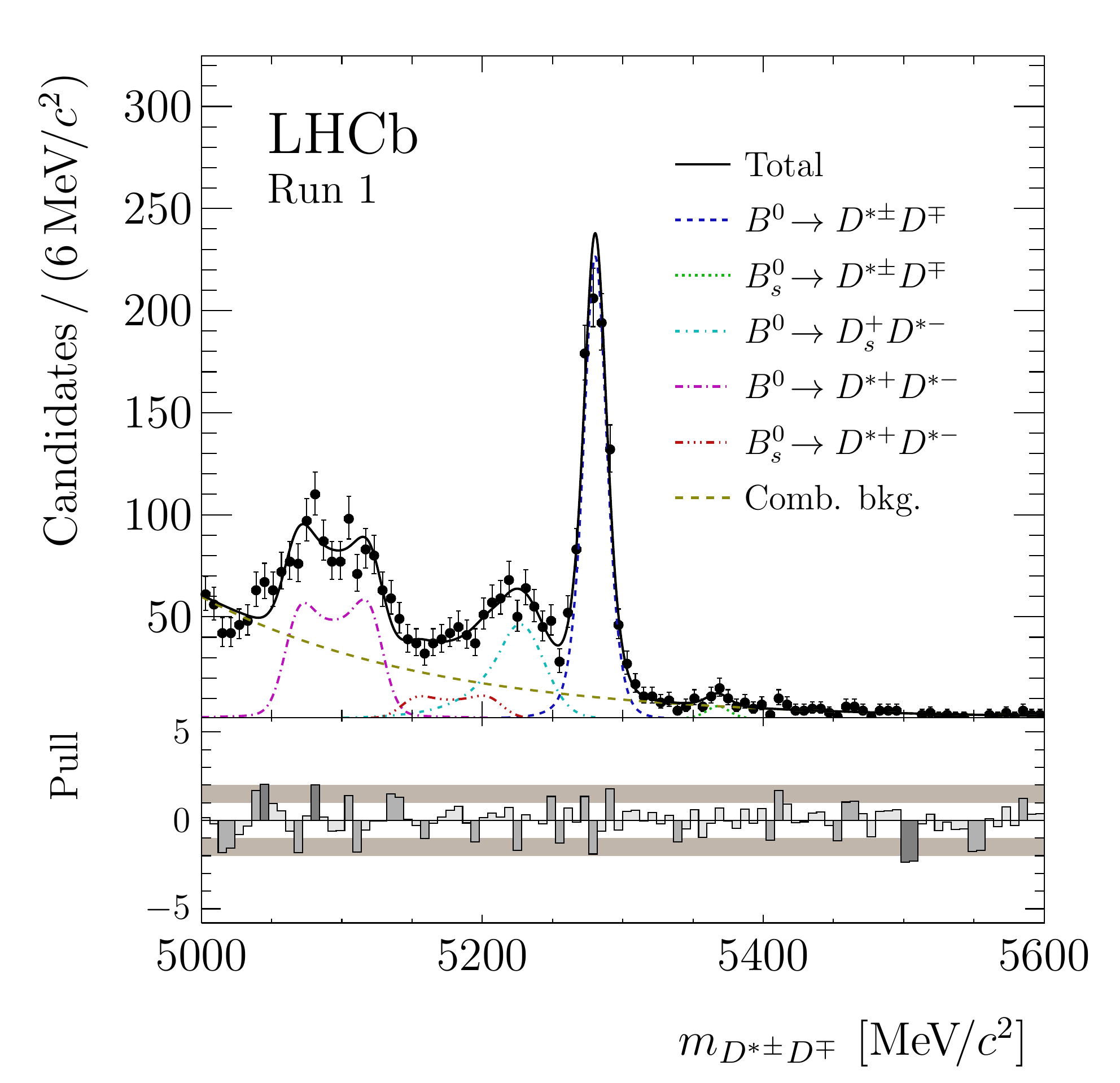}\\ 
\includegraphics[width=0.49\textwidth]{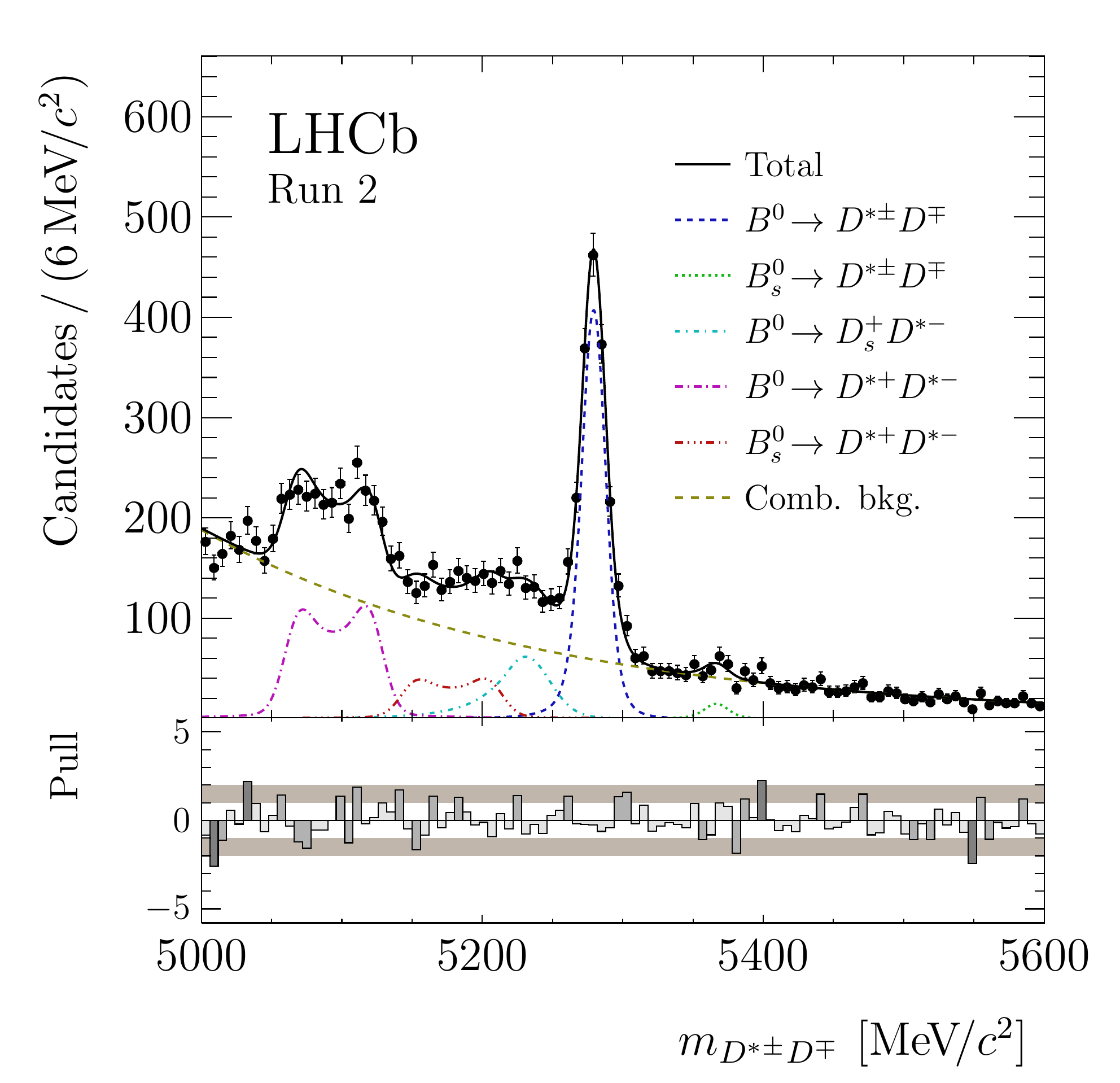}
\includegraphics[width=0.49\textwidth]{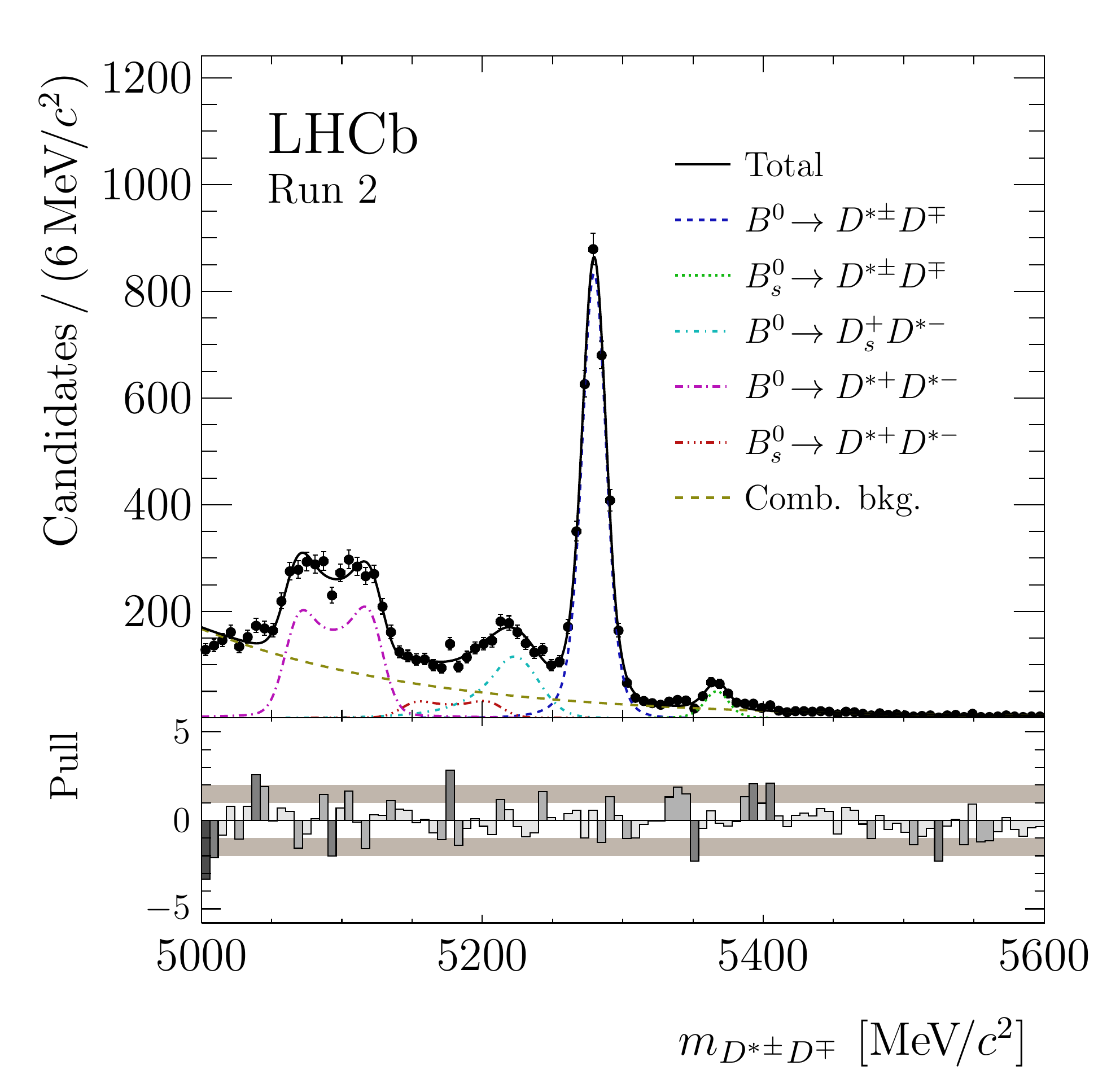}
\caption{ Mass distributions for the \BdToDstD decay with (left) \DzToKpipipi and (right) \DzToKpi for (top) Run 1 and (bottom) Run 2  data samples. 
Besides the data points and the full \PDF (solid black) the projections of the \Bd signal (dashed blue), the \BsToDstD background (dotted green), the \BdToDstDs background (dash-dotted turquoise), the \BdToDstDst background (long-dash-dotted magenta), the \BsToDstDst background (dash-three-dotted red) and the combinatorial background (long-dashed green) are shown.}
\label{fig:widemassfitk3pikpi}
\end{figure}

An unbinned extended maximum-likelihood fit is performed on the invariant-mass distribution of the selected \Bz candidates, on each of the four data samples independently. The \sPlot technique~\cite{Pivk:2004ty} is employed to determine per-candidate weights that are used for background subtraction in the subsequent decay-time fit.
The model describing the \DstD mass distribution consists, in addition to the signal \BdToDstD,  of the following background components: \Bs decays to the same  \DstD final states; 
\BdToDstDs decays with a misidentified kaon that pass the selection; \Bz and \Bs decays to  \mbox{\Dstarp\Dstarm} with one of the excited charm mesons decaying into a charged \D meson and an additional unreconstructed neutral pion or photon; 
and combinatorial background. 
The distribution of the reconstructed mass of the \Bd signal component is parametrised with the sum of two (three) Crystal Ball functions~\cite{Skwarnicki:1986xj} for the \Kpipipi (\Kpi) \Dz final state, with common mean 
but different width and tail parameters. 
The mass model for the \Bs decays is the same as for the \Bd decays, but the peak position is shifted by the difference between the known values of the \Bs and \Bd mesons~\cite{PDG2018}.
The mass distribution of \BdToDstDs decays is described with the sum of two Crystal Ball functions with a common mean, which is floated in the data fit, while the remaining parameters are fixed from simulation.
The mass distribution for the partially reconstructed \BdToDstDst and \BsToDstDst background contributions are described by a combination of functions corresponding to pure longitudinal and transverse polarizations of the two \Dstarpm mesons. 
The relative fraction of the two possible contributions are floated in the fit, while the shapes, that are a double peak in case of longitudinal polarization and a single broad one for transverse polarization, are fixed to those evaluated on simulated samples. 
The relative fraction of the two contributions in \BsToDstDst and \BdToDstDst decays are assumed to be the same.
The distribution of the reconstructed mass for the combinatorial background component is modelled by an exponential function.
The results of the fits to the four data samples are shown in Fig.~\ref{fig:widemassfitk3pikpi} with the partial contribution of each component overlaid; 
the resulting signal yields are $469\pm28$ ($1570\pm48$) and  $856\pm32$ ($3265\pm61$) for the \Dz final states \Kpipipi and \Kpi in Run 1 (Run 2), respectively.

\section{Flavour tagging}
\label{sec:Tagging}
Measurements of \CP violation in decay-time-dependent analyses of \Bz meson decays require the determination of the production flavour of the
\Bz meson. Methods to infer the initial flavour of a reconstructed
candidate, \ie whether it contained a $\bquark$ or a $\bquarkbar$ quark at
production, are referred to as flavour tagging. 
Two classes of algorithms are used. The opposite side (OS) tagger exploits the
fact that $b$ and $\overline{b}$ quarks are almost exclusively produced in pairs
in $pp$ collisions, allowing the
flavour of the signal \Bd candidate to be inferred from the flavour of the other
$b$ hadron in the event. The OS tagger combines information on the charge of the
muon or electron from semileptonic $b$ decays, the charge of the kaon from the
$b\to c\to s$ decay chain, the charge of a reconstructed secondary charm hadron and the charges of the tracks
that form the secondary vertex of the other \mbox{$b$-hadron} decay, combined into a weighted average, with weights depending on the transverse momenta of the tracks \cite{LHCb-PAPER-2011-027,LHCb-PAPER-2015-027}. The same-side (SS) tagger exploits the production of correlated protons or pions in the hadronization of the \bquarkbar (\bquark) quark that forms the signal
\Bd (\Bdb) candidate, with its initial flavour identified by the charge of the particle~\cite{LHCb-PAPER-2016-039}. 

Each tagging algorithm provides a flavour-tagging decision, $d$,
and an estimate, $\eta$, of the probability that the decision is incorrect
(mistag) for a reconstructed \Bd candidate.
The tagging decision takes the value $\pm 1$ for tagged \B candidates
and 0 if no decision on initial flavour can be
assigned (untagged). The mistag probability varies in the range from 0 to 0.5 for tagged candidates and is equal to 0.5 for untagged candidates.

Each tagging algorithm is implemented as a BDT that is trained and
optimised using large data samples 
of $\Bu\to\jpsi\Kp$ decays for the OS and of $\Bd\to\Dm\pip$ decays for the SS taggers, respectively~\cite{Fazzini:2018dyq}.
The mistag probability for each tagger is given by the
output of the BDT, which is calibrated using dedicated data control channels to
relate $\eta$ to the true mistag probability, $\omega$. The performance of the flavour tagging is measured by the tagging power, \effD, where
\etag is the fraction of tagged candidates and $D=1-2\omega$ represents the dilution induced on the oscillation amplitude. 
The tagging power represents the effective loss in signal yield compared to a perfectly tagged sample. 

\subsection{Calibration of the tagging output}
Flavour tagging algorithms are calibrated using control samples of flavour-specific \B decays, separately in Run 1 and Run 2 data.
At first approximation, the measured mistag fraction $\omega$ in the control channel can be expressed as a linear function of the predicted mistag estimate $\eta$ as
\begin{equation}
\label{eq:tagging:calib_function}
  \omega(\eta) = p_0 + p_1\left(\eta - \langle\eta\rangle\right) ,
\end{equation}
where the use of the arithmetic mean $\langle\eta\rangle$ of the $\eta$ distribution decorrelates the $p_0$ and $p_1$ parameters. A perfect calibration of the taggers would result in $p_0 = \langle\eta\rangle$ and $p_1 = 1$. 
A signal candidate tagged with decision $d$ and a calibrated mistag $\omega > 0.5$ corresponds to an opposite tagging decision $d' = - d$ with a mistag probability  $\omega' = 1 - \omega$.

The performance of the flavour taggers, \etag and \mistag, may depend on the
initial flavour of the neutral \B meson. Charged decay products, like the \Kp and \Km
which are used by the OS kaon tagger, can have significantly different
interaction rates with the detector material and therefore lead to different reconstruction efficiencies. 
To account for these asymmetries Eq.~\ref{eq:tagging:calib_function} is modified as 
\begin{eqnarray}
\label{eqn:tagging_calib}
{\kern \thebaroffset\optbar{\kern -\thebaroffset \omega}{}} (\eta) &=& \left( p_0  \pm \frac{\delta p_0}{2}\right) +
\left(p_1 \pm \frac{\delta p_1 }{2}\right) \left(\eta - \langle\eta\rangle\right),
\end{eqnarray} 
where the mistag fractions $\omega$ and $\overline{\omega}$ for an initial \Bz and \Bzb correspond to the plus and minus sign, respectively, and $\delta p_{0,1}= p_{0,1}^{\Bz}- p_{0,1}^{\Bzb}$.
Similarly, the tagging asymmetry, ${\cal A}_{\rm tag}$, is defined as
\begin{equation}\label{eq:Atag}
{\cal A}_{\rm tag}=\frac{\varepsilon_{\rm tag}^{\Bdb}-\varepsilon_{\rm tag}^{\Bd}}{\varepsilon_{\rm tag}^{\Bdb}+\varepsilon_{\rm tag}^{\Bd}}\, .
\end{equation}

For this analysis, the calibration of the OS and SS taggers is performed  using samples of \BdToDsD and \BdToDstDs flavour-specific decays which have similar topology, kinematics and event characteristics as the signal decay.
Candidate \Bd mesons in the two decay modes are reconstructed using the decays
\decay{\Dsp}{\Km\Kp\pip}, \mbox{\decay{\Dm}{\Kp\pim\pim}} and
\mbox{\decay{\Dstarm}{\pim\Dzb}}, followed by
\mbox{\decay{\Dzb}{\Kp\pim}} or
\mbox{\decay{\Dzb}{\Kp\pim\pip\pim}}. 
These decays are selected using the same trigger requirements, and similar kinematic and geometric criteria, to those applied for the signal selection.
Requirements are applied on the identification of the final-state particles as well as on the invariant mass of \Ds, \Dstarm and \Dzb mesons. Candidate \Bz mesons are retained if their invariant mass and decay time are in the ranges $5220-5500\mevcc$ and $0.3-10.3\ps$, respectively. In case multiple candidates are selected in the same event, the candidate with the highest \pt is retained.

A fit to the invariant-mass distribution of the \Bd candidates is performed on each sample independently to separate the signal from the background contribution, which consists of random combination of particles. The signal is modelled by the sum of two Gaussian functions with a common mean, while the background shape is described by an exponential function.
In the case of the \BdToDsD control channel, an additional contribution due to \BsToDsD decays  is also needed. It is modelled by the same function used for the \Bd decays except for the mean value, that is shifted by the known difference of the \Bs and \Bd masses~\cite{PDG2018}.
Signal yields of 11,400 (39,200) \BdToDstDs decays and 24,900 (102,900)  \BdToDsD decays, in the Run 1 (Run 2) data sample, are found. 

The mass fit determines for each selected candidate a weight that is used to statistically subtract the background contribution, using the \sPlot technique. 
The tagging calibration parameters are determined from an unbinned maximum-likelihood fit to the weighted distribution of the decay time $t$, the final state $r$, with $r=+1$ $(-1)$ for $D_s^{+}D^{(\ast)-}$ ($D_s^{-}D^{(\ast)+}$), 
tag decisions $\vec{d}=(d_{\rm OS}, d_{\rm SS})$ and probabilities $\vec{\eta}=(\eta_{\rm OS}, \eta_{\rm SS})$, according to the following \PDF
\begin{equation}
  {\cal P} \left(t,r,\vec{d} \, \lvert \, \vec{\eta} \right) = \epsilon(t) \cdot \left[ \mathcal{F}(t',r,\vec{d} \, \lvert \, \vec{\eta}) \otimes  {\cal R}(t-t^\prime) \right] \, .
\end{equation}
Here, ${\cal R}(t-t^\prime)$ is the decay-time resolution, which is modelled using simulation.
The decay-time efficiency, $\epsilon(t)$, is parametrised using a cubic spline model~\cite{GRnote}, with parameters determined from the fit to the data.
The number of spline coefficients and the knot positions are very similar among the \BdToDsD and \BdToDstDs decays, as well as among Run 1 and Run 2 samples.
The signal \PDF, $\cal F$, describes the decay-time distribution of flavour-specific \Bz decays; it is based on 
Eqs.~\ref{eq:timepdfdstd2} and~\ref{eq:timepdfdstd7} of Sec.~\ref{sec:Decaytimefit}, with appropriate values for the coefficients corresponding to flavour-specific \Bd decays.
It depends on the $\tau_d$ and \dmd parameters, that describe the \Bd decay and mixing, which are fixed to their known values~\cite{PDG2018}. 
In addition, the \PDF depends on the tagging parameters of Eq.~\ref{eqn:tagging_calib}, on the tagging efficiencies and their asymmetry,  on the \Bd production asymmetry and on the detection asymmetry of the  control channel, that are determined from the fit to data.
The asymmetry of the \Bz meson production is defined as 
${\cal A}_{\mathrm{prod}}  = {(\sigma_{\Bzb} - \sigma_{\Bz})}/{(\sigma_{\Bzb} + \sigma_{\Bz})}$,
where $\sigma_{\Bzb}$ and $\sigma_{\Bz}$ are the production rates of \Bzb and \Bz mesons.
 
The fit is performed independently on each sample. For each data-taking period, the parameters of the tagging calibration and of the production asymmetries are consistent between the two channels,
and the results are combined according to their statistical uncertainties. 

Different sources of systematic uncertainty are considered. 
Systematic uncertainties  on the calibration parameters are determined by repeating the fit to data with modified assumptions. The deviations of the fit results from the nominal values are assigned as systematic uncertainties.
The largest uncertainty is related to the determination of the signal weights from the fit to the mass distribution. An alternative determination of such weights is performed fitting the two-dimensional invariant-mass distribution of the \Ds and \Dzb (\Dstarm) candidates. 
To account for possible effects related to the decay-time efficiency model, an analytic function is considered instead of the cubic spline function. 
The uncertainties related to the input values of the \Bd decay and mixing properties are determined repeating the fit with inputs varied within their uncertainties.
Finally, uncertainties related to differences between data and simulation concerning the time resolution model are neglected given their insignificant impact on the tagging parameters for variations up to 50\%. The resulting tagging parameters are listed in Table~\ref{tab:taggingcalibration}.
\begin{table}[!b]
\begin{center}
\caption{Flavour-tagging parameters obtained as a weighted average of the values measured in the two control channels.
The quoted uncertainties are statistical and systematic.\label{tab:taggingcalibration}}
\renewcommand{\arraystretch}{1.15}
\begin{tabular}{l r@{$\,\pm\,$}l r@{$\,\pm\,$}l }
\hline
Parameter 			& \multicolumn{2}{c}{Run 1}     & \multicolumn{2}{c}{Run 2} \\
\hline
${\cal A}_{\rm prod}$      & $-0.011$ & 
$0.008\pm0.003$ &  0.004 & $0.005\pm0.002$ \\
\hline
${\cal A}_{\rm tag}^{\rm OS}$ 
                    &   0.008 & $0.008\pm 0.005$  &  0.001 & $0.004\pm0.009$ \\
$\delta p_0^{\rm OS}$	&   0.022 & $0.007\pm0.013$   & 0.009 & $0.004\pm0.003$  \\
$\delta p_1^{\rm OS}$	&   0.20   & $0.06\phantom{0}\pm0.11$   & 0.02  & $0.03\phantom{0}\pm0.01$\\
$p_0^{\rm OS} -\langle \eta^{\text{OS}}\rangle $
					&  0.037 & $0.005\pm0.006$  & 0.032 & $0.003\pm0.005$ \\
$p_1^{\rm OS}$		&  0.95 & $0.04\phantom{0}\pm0.10$    & 0.87 & $0.02\phantom{0}\pm0.02$ \\
\hline
${\cal A}_{\rm tag}^{\rm SS}$ 
					    &  $-0.006$ & $0.004\pm0.005$ & $-0.001 $  & $0.001\pm0.002$  \\
$\delta p_0^{\rm SS}$	&  $-0.004$ & $0.005\pm0.009$ & 0.001      & $0.003\pm0.003$ \\
$\delta p_1^{\rm SS}$	&  $-0.03$ & $0.08\phantom{0}\pm 0.06$ & $-0.11$  & $0.04\phantom{0}\pm 0.02$ \\ 
$p_0^{\rm SS} - \langle \eta^{\text{SS}}\rangle $
					    &  0.007 & $0.004\pm0.006$  & $-0.002$ & $0.002\pm0.002$ \\
$p_1^{\rm SS}$			&  0.94 & $0.06\phantom{0}\pm 0.07$    & 0.88 & $0.03\phantom{0}\pm0.03$ \\
\hline
\end{tabular}
\end{center}
\end{table}

The portability of the calibration from the control channels to the signal decay channel is assessed with simulation. Both OS and SS taggers show compatible results
among the different decay modes. A deviation from linearity of the SS tagger calibration is observed in the signal channel. Its impact on the determination of the signal \CP parameters is evaluated using pseudoexperiments where the effect is reproduced at generation level, as described in Sec.~\ref{sec:Systematics}.
In addition, a dependence of the OS tagging efficiency on the decay-time for $t<1$\ps is present in data. Pseudoexperiments are used to determine the related uncertainties on the \CP parameters, as described in Sec.~\ref{sec:Systematics}.

\subsection{Tagging results}
Approximately 40\% of the tagged candidates in the signal decay samples are tagged by both the OS and the SS algorithms. 
Since the algorithms select different samples of charged particles and hence are uncorrelated, the two tagging results are combined taking into account both decisions and their corresponding estimates of $\eta$ as detailed in Ref.~\cite{LHCB-PAPER-2018-009}. 
The combined estimated mistag probability and the corresponding uncertainty are obtained by combining the individual calibrations for the OS and SS taggers and
propagating their uncertainties in the decay-time fit. 
The effective tagging power and efficiency 
for signal candidates tagged by one or both of the OS and SS algorithms are given in Table~\ref{tab:taggingPerformances}.

\begin{table}[t]
\centering
\caption{
Tagging efficiency and tagging power for \BdToDstD signal candidates in the four data samples, computed using the event-by-event predicted mistag $\eta$ and the calibration parameters obtained from control channels. The quoted uncertainties are statistical only.}
\renewcommand{\arraystretch}{1.15}
\begin{tabular}{ll r@{$\,\pm\,$}l  r@{$\,\pm\,$}l @{$\,\pm\,$}l r@{$\,\pm\,$}l  r@{$\,\pm\,$}l @{$\,\pm\,$}l}
 \hline  
       & & \multicolumn{5}{c}{Run 1} & \multicolumn{5}{c}{Run 2} \\
Sample &Tagger& \multicolumn{2}{c}{\etag [$\%$] } & \multicolumn{3}{c}{\effD [$\%$]} & \multicolumn{2}{c}{\etag [$\%$] } & \multicolumn{3}{c}{\effD [$\%$]}\\
\hline
\multirow{4}{*}{\DzToKpipipi}
&OS only    & 8.3 & 1.6 & 0.64 & \multicolumn{2}{l}{\kern-0.5em 0.18}  
            & 3.9 & 0.6 & 0.36 & \multicolumn{2}{l}{\kern-0.5em 0.08}  \\
&SS only    & 43.0 & 2.9  & 1.17 & \multicolumn{2}{l}{\kern-0.5em 0.16}  
            & 47.4 & 1.5  & 1.57 & \multicolumn{2}{l}{\kern-0.5em 0.11}  \\
&OS\&SS both  & 37.5 & 2.9  & 4.44 & \multicolumn{2}{l}{\kern-0.5em 0.57}  
            &  41.5 & 1.5  & 5.11 & \multicolumn{2}{l}{\kern-0.5em 0.30}  \\
\cline{2-12}
&Total      & 88.8 & 1.9  & 6.25 & \multicolumn{2}{l}{\kern-0.5em 0.55}   
            & 92.7 & 0.8  & 7.05 & \multicolumn{2}{l}{\kern-0.5em 0.29}  \\
\hline
\multirow{4}{*}{\DzToKpi}
&OS only    & 12.2 & 1.2  & 1.14 & \multicolumn{2}{l}{\kern-0.5em 0.19}  
            &  4.2 & 0.4 & 0.42 & \multicolumn{2}{l}{\kern-0.5em 0.06}  \\
&SS only    & 40.3 & 1.8  & 1.43 & \multicolumn{2}{l}{\kern-0.5em 0.18}  
            & 51.4 & 0.9  & 1.61 & \multicolumn{2}{l}{\kern-0.5em 0.07}  \\
&OS\&SS both &  27.7 & 1.7  & 3.05 & \multicolumn{2}{l}{\kern-0.5em 0.30}
            & 37.9 & 0.9  & 4.57 & \multicolumn{2}{l}{\kern-0.5em 0.19}  \\
\cline{2-12}
&Total      & 80.2 & 1.4  & 5.61 & \multicolumn{2}{l}{\kern-0.5em 0.36} 
            & 93.5 & 0.5  & 6.61 & \multicolumn{2}{l}{\kern-0.5em 0.19}   \\
            \hline
\end{tabular}
\label{tab:taggingPerformances}
\end{table}	
\section{Instrumental asymmetries}
\label{sec:Asymmetries}
The overall asymmetry that is measured in the decay-time fit has to be corrected for instrumental asymmetries in order to determine the physical parameter \ADstD. These asymmetries affect reconstruction, detection and particle identification efficiencies and are related to the different interaction cross-section with matter and different detection and identification efficiencies of positive and negative pions and kaons. The \BdToDstD decay is charge symmetric, however since all instrumental efficiencies depend on momenta, and the $p$ and \pt spectra of kaons and pions in the \Dstarpm and \Dmp decays are observed to be slightly different, the cancellation is not expected to be complete. 

To a good level of approximation, the asymmetry in \BdToDstD decays, denoted with $\Adet$, can be related to the  \Dstarm and \Dm  asymmetries, $\Adet^{\Dstarm}$ and $\Adet^{\Dm}$, as
\begin{align}
\Adet^{D^{*-}D^+} &\simeq \Adet^{D^{*-}} - \Adet^{D^{-} }\, .
\end{align}
Each of the \Dstarm and \Dm asymmetries is measured using a sample of prompt \DToKpipi decays after a kinematic weighting is applied to match the distribution of the final state particles of the signal, as described in the following. 
In order to account also for the PID asymmetry, the same kaon and pion identification requirements used in the \BdToDstD  selection are applied to the prompt \Dm meson sample.  The additional asymmetries induced by the use of PID variables in vetoes and in the BDT employed for the \BdToDstD selection are considered as a source of systematic uncertainty on the asymmetry $\Adet^{D^{*-}D^+}$.
A sample of about 6~million prompt \DToKpipi decays collected  in Run 2 is used and divided into four subsamples: for each \Dz decay mode of the \BdToDstD signal, one sample is used to measure the \Dstarm asymmetry, the other is used to determine the \Dm asymmetry.
In each subsample a fit to the $\Kp\pim\pim$ invariant mass distribution is performed to determine the weights to be used to subtract background. The signal mass model is parametrised by two Crystal Ball functions with common mean and the background is modelled by an exponential function.  
For the \DzToKpi decay mode, one of the prompt \DToKpipi subsamples is weighted to match the spectra of the final state particles in the signal \Dstarp decay 
and the other is weighted to match the spectra of the final state particles in the signal \Dm decay~\cite{Rogozhnikov:2016bdp}. 
The same procedure is used for the \DzToKpipipi decay mode, but not considering two of the pions with opposite charge from the \Dz decay, since the $\pi^-\pi^+$ pair is found to contribute negligibly to additional asymmetries. 
Finally, the asymmetries  $\Adet^{D^{*-}}$ and  $\Adet^{D^{-} }$ are calculated from the \Dp and \Dm yields obtained by fitting the mass distributions of each weighted prompt \Dp and \Dm meson samples and subtracted.
The results are reported in Table~\ref{tab:asymmetries} and $\Adet^{\Dstarm \Dp}$ values are consistent with zero.

\begin{table}[!tb]
	\centering
	\caption{Instrumental asymmetries of \Dpm, \Dstarpm and their combination for \BdToDstD decays obtained from prompt \D meson decays. The quoted uncertainties are statistical, in the last column statistical and systematic.}
	\label{tab:asymmetries}
	\renewcommand{\arraystretch}{1.15}
	\begin{tabular}{l r@{$\,\pm\,$}l r@{$\,\pm\,$}l | r@{$\,\pm\,$}l}
		\hline
		Final state		& \multicolumn{2}{c}{$\Adet^{\Dstarm}$} &\multicolumn{2}{c}{$\Adet^{\Dm}$ } & \multicolumn{2}{c}{$\Adet^{\Dstarm \Dp}$}\\
		\hline 
		\DzToKpi      & 0.0169 & 0.0036 & 0.0158 & 0.0018 & 0.0011 & $0.0040\pm 0.0030$ \\ 
		\DzToKpipipi  & 0.0146 & 0.0022 & 0.0138 & 0.0015 & 0.0009 & $0.0026\pm 0.0051 $ \\ 
		\hline
	\end{tabular}
\end{table}
The following sources of systematic uncertainties on \Adet are considered:  asymmetries due to the particle identification and hadronic hardware-trigger requirements,  variation of the prompt \DToKpipi selection criteria  and of the mass-fit model and imperfect cancellation of the prompt \Dpm production asymmetry. 
Since the differences between the kinematic distributions of the final state particles of \Dstarpm and $\Dmp$ decays are observed to be smaller in Run 1 than in Run 2,  the difference of \Dstarpm and \Dmp asymmetries in Run 1 is assumed to be zero with the same statistical and systematic uncertainties as in Run 2.

\section{Decay-time fit} \label{sec:Decaytimefit}
A decay-time fit is performed simultaneously on the four data samples to measure the \CP coefficients \SDstD, \DeltaSDstD, \CDstD, \DeltaCDstD and \ADstD.
The weights determined with the mass fit are used to subtract the background,  using the \sPlot technique. 
Before performing the fit, a blinding transformation is applied on the \CP parameters that is removed only at the end of the analysis.

The \PDF describing the measured \Bz decay-time distribution and tag decisions $\vec{d}=(d_{\rm OS}, d_{\rm SS})$, given the mistag
probability estimates $\vec{\eta}=(\eta_{\rm OS}, \eta_{\rm SS})$, can be expressed as 
\begin{equation}
\label{eq:timepdfdstd}
{\cal P}\left(t,r,\vec{d} \, | \, \vec{\eta}\right) = \epsilon(t)\cdot \Bigl( {{\cal F}}(t',r,\vec{d} \, | \, \vec{\eta}) \otimes {\cal R}(t-t') 
\Bigr) ,
\end{equation}
where ${\cal P}(t',r,\vec{d}\, |\, \vec{\eta})$ is the \PDF describing the distribution of the true decay time, $t'$, ${\cal R}(t-t')$
is the decay-time resolution function, while $\epsilon(t)$ describes the decay-time efficiency.
The \PDF describing the \Bz decay-time distribution has the same form as Eq.~\ref{eq:cpvdstd5} but with effective coefficients 
\begin{equation}
\label{eq:timepdfdstd2}
{{\cal F}}(t' ,r ,\vec{d} \, | \, \vec{\eta}) = N {e}^{-t'/ \tau_d} (1 + r  \Araw) \Bigl[ C^{\text{eff}}_{\cosh}- C^{\text{eff}}_{\sin} \sin\Bigl(\dmd t' \Bigr) + C^{\text{eff}}_{\cos} \cos \Bigl(\dmd t'\Bigr) \Bigr] .
\end{equation}
These coefficients, that depend on the final state variable $r=\pm 1$, can be expressed as
\begin{equation}
\label{eq:timepdfdstd7}
\begin{split}
C^{\text{eff}}_{\sin} &=
(\SDstD+r\DeltaSDstD)(\Delta^- - {\cal A}_{\text{prod}}\Delta^+),\\ 
C^{\text{eff}}_{\cos} &=
(\CDstD+r\DeltaCDstD)(\Delta^- - {\cal A}_{\text{prod}}\Delta^+),\\
C^{\text{eff}}_{\cosh} &= (\Delta^+ - {\cal A}_{\text{prod}}\Delta^-), 
\end{split}
\end{equation}
where the factors $\Delta^{\pm}$ contain the dependence on the mistag fraction, the tagging decisions and efficiencies and the asymmetries in the tagging efficiencies of the OS and SS taggers. 
The parameter \Araw is the sum of the \CP asymmetry, \ADstD, and the detection asymmetry, $\Adet^{\Dstarm\Dp}$. Since the instrumental asymmetries depend on the final state and on the data-taking period, four different parameters are used in the fit.

The decay-time resolution model is determined from simulation and fixed in the fit to data with an effective resolution of {$60\fs$}.
The same set of parameters describes well both Run~1 and Run~2 data, with small differences between the two \Dz final-state samples. 
Due to the low \Bz--\Bzb oscillation frequency, the decay-time resolution has a very small influence on the relatively low \CP parameters.

The selection and reconstruction efficiency depends on the \Bz decay time due to displacement requirements made on the signal final-state particles and a decrease in the reconstruction efficiency for tracks with large impact parameter with respect to the beam line~\cite{LHCB-PAPER-2013-065}. 
The decay-time efficiency is modelled by a cubic spline function with five knots~\cite{GRnote}. 
The knot positions are chosen from a fit to simulated signal candidates at (0.3, 0.5, 2.7, 6.3, 10.3)\ps. 
The spline coefficients are determined from the fit to data where the \Bd lifetime is fixed to its known value~\cite{PDG2018}. 
The same parameters are assigned to both \Kpi, \Kpipipi final-state samples, while different values are used for Run~1 and Run~2 data, as suggested by simulation studies. 
The mass difference $\Delta m_d$ is fixed to its known value~\cite{PDG2018}.

The average mistag values for OS and SS tags are fixed to the values calculated in each corresponding signal sample.
All the  tagging parameters, as reported in Table~\ref{tab:taggingcalibration}, are introduced in the fit through Gaussian constraints, in order to account for their associated uncertainties. The tagging efficiencies are free to vary in the fit.
The OS and SS taggers are combined in the fit.
The production asymmetries are also constrained in the fit to the values measured in the flavour-tagging control channels. 
The result of the production asymmetry in Run~1 is in agreement with the dedicated \lhcb measurement~\cite{LHCb-PAPER-2016-062}, when considering the kinematics of the signal data sample. 

The \CP observables resulting from the decay-time fit are
\begin{equation*}
\label{eq:dstdresults}
\begin{split}
    \SDstD                	&= -0.861\pm0.077,\\
    \DeltaSDstD             &= \phantom{+}0.019\pm0.075, \\
    \CDstD                	&= -0.059\pm0.092, \\
    \DeltaCDstD          	&= -0.031\pm0.092. \\
\end{split}
\end{equation*}
After subtracting the detection asymmetries to each of the four \Araw values the weighted mean is calculated as \mbox{$\ADstD=0.008 \pm 0.014$}.
The quoted uncertainties include contributions due to the size of the samples and due to the external parameters constrained in the fit. 
The correlations among the \CP parameters are 
\begin{equation*} \label{eq:dstdfitcorrelation}
\centering
 \begin{array}{c|c c c c}
                 & \SDstD & \DeltaSDstD  &\CDstD & \DeltaCDstD\\ 
         \hline
         \SDstD        & 1    &  0.07  & 0.44    & 0.05 \\
         \DeltaSDstD   &      &   1    & 0.04    & 0.46 \\
         \CDstD        &      &        &  1      & 0.04 \\
         \DeltaCDstD   &      &        &         &  1   \\
        
      \end{array}\, .
\end{equation*}
The four \Araw parameters are almost uncorrelated among each other, with  
the largest correlation coefficient being $10^{-4}$. 
Their correlation with all the other fit parameters is also small, between 0.1\% and 1\%. The spline parameters have large correlations among them but have correlations between 0.1\% and 1\%  with the \CP parameters. 

\begin{figure}[t]
\centering
\includegraphics[width=0.6\textwidth]{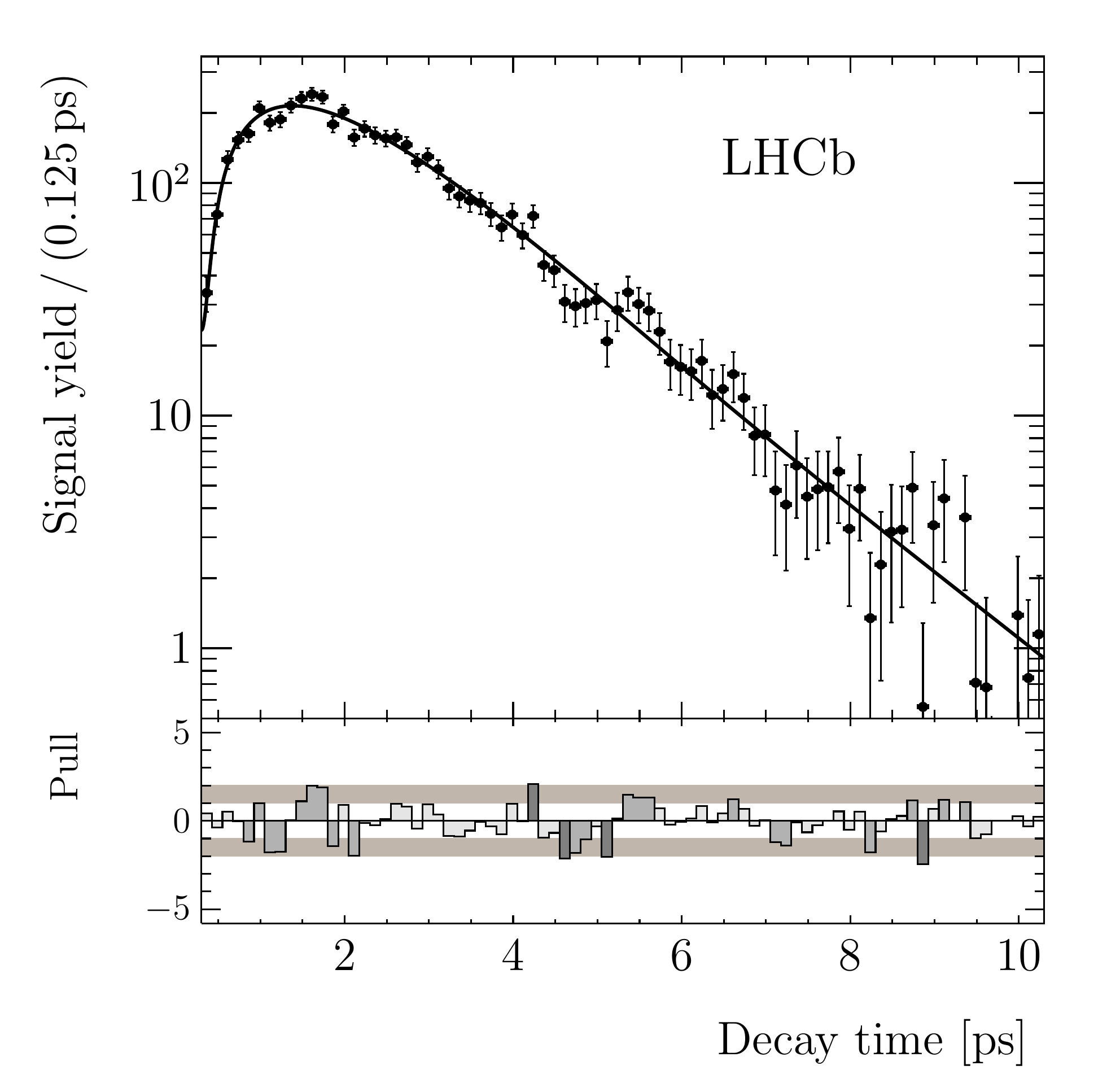} 
\caption{Decay-time distribution of the \BdToDstD signal candidates, summed over all data samples, where the background contribution is subtracted by means of the \sPlot technique. The projection of the \PDF is represented by the full line.}
\label{fig:dstddecaytimefit}
\end{figure}

\begin{figure}[t]
\setlength{\unitlength}{1mm}
\centering
\includegraphics*[width=0.75\textwidth]{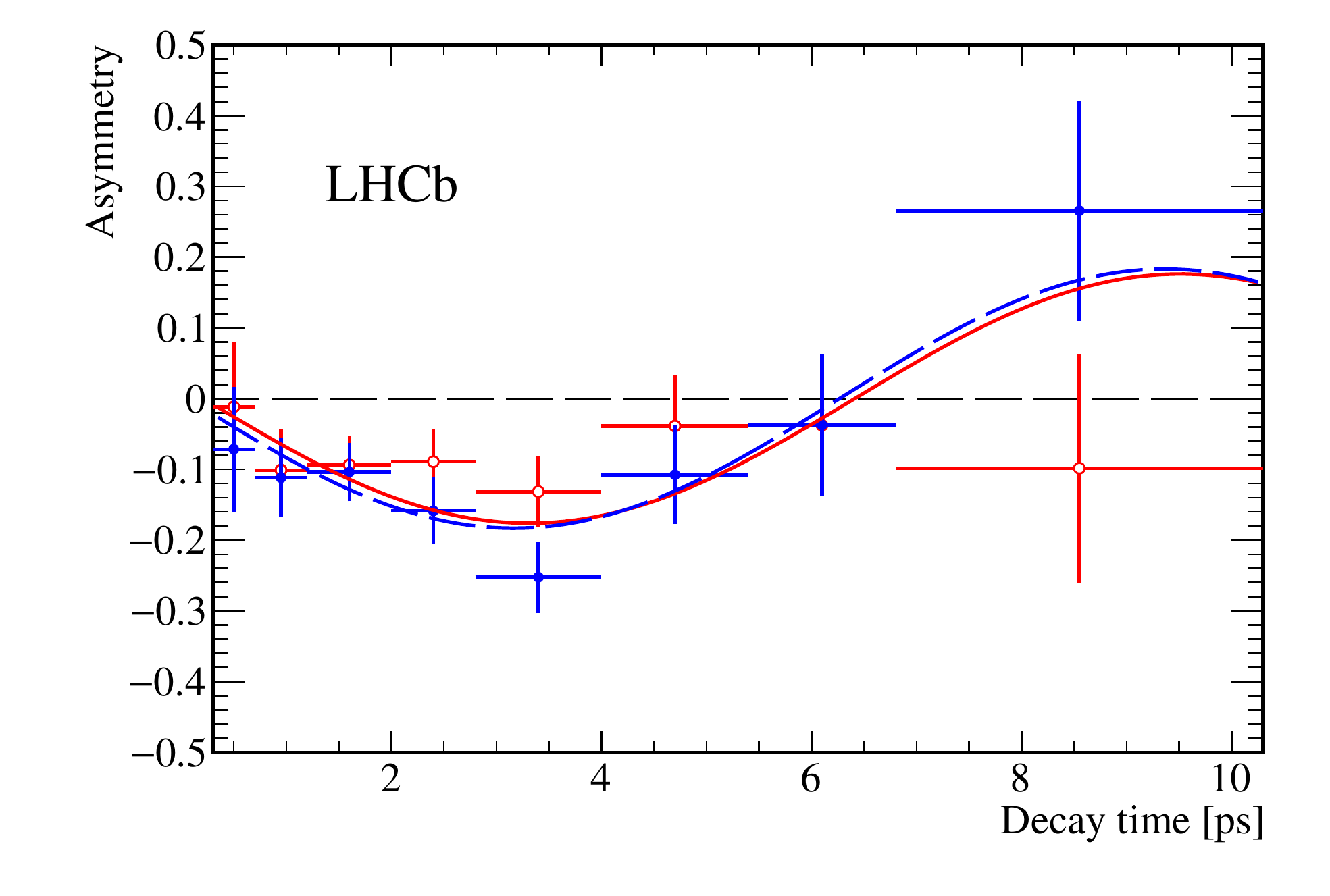}
\caption{ Asymmetry between \Bzb and \Bz signal yields as a function of the decay time, for (blue full dot) \Dstarp\Dm and (red empty dot) \Dstarm\Dp signal candidates with a non null tagging decision. The background contribution is subtracted by means of the \sPlot technique. The corresponding projections of the \PDF are represented by the blue dashed (red continuous) line. 
The \Bd flavour is determined by the combination of all flavour-tagging algorithms.}
\label{fig:CPasymmetry}
\end{figure}

Figure~\ref{fig:dstddecaytimefit} shows the decay-time distribution of the full \BdToDstD data sample, where the  fitted \PDF is overlaid.
For illustration, Fig.~\ref{fig:CPasymmetry} shows the asymmetry between \Bzb and \Bz signal yields as a function of the decay time, separately for  \Dstarp\Dm and \Dstarm\Dp final states, with the corresponding fit functions overlaid.


\section{Systematic uncertainties}
\label{sec:Systematics}
Several cross checks of the analysis and possible sources of systematic uncertainties of the results are considered in the following and summarised in Table~\ref{tab:syst_total}.
As a first validation, the decay-time fit is performed on a simulated sample of signal decays corresponding to 36 times the data sample size. 
The resulting \CP parameters are compatible with the generation values within less than two standard deviations. 
In order to test if the likelihood estimate for the \CP parameter values are accurate, the same mass and decay-time models as used for data are fitted to 2000 pseudoexperiments.
For each pseudoexperiment, four samples are generated corresponding to the four data samples. In each sample, 
the input parameters for the signal component are the same as in the fit to data, except for the production and tagging asymmetries, which are set to zero for convenience.  The \CP parameters, the tagging efficiencies and the coefficients of the splines describing the time acceptance are taken from the best fit result to data. All the parameters that are constrained in the fit are set to values randomly generated according to the constraints applied. Each background contribution is generated with a specific time dependence, which in some cases accounts for \CP violation. 
No bias is found for the \CP parameters nor for the raw asymmetries \Araw. The uncertainty on the mean value of the bias is taken as systematic uncertainty on the parameter.

\begin{table}[t]
\caption{Summary of the systematic uncertainties. The total systematic uncertainties are computed as quadratic sum of individual contributions. }
\label{tab:syst_total}
\centering
	\renewcommand{\arraystretch}{1.15}
    \begin{tabular}{lcccc}
    Source & {$\phantom<$}${\DeltaCDstD}$ & {$\phantom<$}${\CDstD}$& {$\phantom<$}${\DeltaSDstD}$ {$\phantom<$}& {$\phantom<$}${\SDstD}$ \\ 
    \hline
    \small{Fit bias}           &  {$\phantom<$}0.002 & {$\phantom<$}0.002 & {$\phantom<$}0.002 & {$\phantom<$}0.002 \\
    \small{Mass model}         &  {$\phantom<$}0.006 & {$\phantom<$}0.014 & {$\phantom<$}0.003 & {$\phantom<$}0.011 \\
    \small{$\dmd,\tau_d,\DGd$} &  {$\phantom<$}0.001 & {$\phantom<$}0.003 & {$\phantom<$}0.001 & {$\phantom<$}0.001 \\
    \small{Decay-time resolution} & $<$0.001  & $<$0.001 & $<$0.001 & $<$0.001 \\
    \small{Decay-time acceptance} & $<$0.001  & $<$0.001 & $<$0.001  & $<$0.001 \\
    \small{Flavour tagging}    & {$\phantom<$}0.015 & {$\phantom<$}0.014 & {$\phantom<$}0.012 & {$\phantom<$}0.015 \\
    \hline
    Total syst. uncertainty         & {$\phantom<$}${0.016}$    & {$\phantom<$}${0.020}$ &{$\phantom<$}${0.012}$&{$\phantom<$}${0.019}$ \\
    \hline
    \\
    Source &{$\phantom<$}${\ArawKpipipiU}$ &{$\phantom<$}${\ArawKpipipiD}$  & {$\phantom<$}${\ArawKpiU}$ & {$\phantom<$}${\ArawKpiD}$  \\ \hline
    \small{Fit bias} & {$\phantom<$}0.0013 & {$\phantom<$}0.0007 & {$\phantom<$}0.0008 & {$\phantom<$}0.0004\\
    \small{Mass model} & {$\phantom<$}0.0025 & {$\phantom<$}0.0024 & {$\phantom<$}0.0021 & {$\phantom<$}0.0016\\
    \small{$\dmd,\tau_d,\DGd$} & {$\phantom<$}0.0003 & {$\phantom<$}0.0002 & {$\phantom<$}0.0002 & {$\phantom<$}0.0001\\
    \small{Decay-time resolution} & {$\phantom<$}0.0002      & {$\phantom<$}0.0001 & {$\phantom<$}0.0001 & {$\phantom<$}0.0001\\
    \small{Decay-time acceptance}& {$\phantom<$}0.0003     & {$\phantom<$}0.0001 & {$\phantom<$}0.0002 & {$\phantom<$}0.0001\\
    \small{Flavour tagging} & {$\phantom<$}0.0001  & {$\phantom<$}0.0001 & {$\phantom<$}0.0001 & {$\phantom<$}0.0001\\ \hline
    Total syst. uncertainty &{$\phantom<$}0.0028 & {$\phantom<$}0.0025 & {$\phantom<$}0.0023 & {$\phantom<$}0.0016\\
    \end{tabular}
\end{table}

In order to cross-check the statistical uncertainty obtained from the fit to data, a bootstrapping procedure is used~\cite{efron:1979}.
In this frequentist model-independent approach a new data sample is generated by drawing candidates from the nominal data sample until the number of candidates matches that of the original one (the same event can be drawn multiple times).
The nominal fit to the decay time is executed, the fit result is stored and the distribution of the residuals with respect to the starting values of the parameters is analysed. 
Given that the standard deviation of the Gaussian fits of the residuals agree with the mean values of the uncertainty from the fit it can be assumed that the uncertainty of the nominal fit is accurate.

Cross checks on the stability of the result are performed by dividing the data sample into categories, according to the \Dz final state, the tagging algorithm (OS and SS) and the magnet polarity. No evidence of bias is found, as all variations of the fit parameters obtained in the splittings are smaller than two standard deviations.   

Specific studies are performed to estimate the effect of an inaccurate determination of the mass model, the decay-time model, the flavour-tagging performance and variations of the input parameters to the decay-time fit.
Four systematic effects on the mass model, described below, 
are considered and the sum in quadrature of each systematic uncertainty is taken as uncertainty on the mass model reported in Table~\ref{tab:syst_total}.
The signal model is changed to two (three) Crystal Ball functions for the mass distribution of \Kpi (\Kpipipi) final state or to two Crystal Ball functions for both final states. The model for the combinatorial background is changed to a second-order polynomial.
These systematic uncertainties are evaluated using pseudoexperiments, with a procedure that is used also in the rest of the paper for other sources of uncertainty.
In each pseudoexperiment, the mass fit to each of the four samples is performed with the nominal model and then repeated with the alternative model. Results of the subsequent decay-time fit are compared to those obtained with the nominal fit and the distribution of their difference is built. The systematic uncertainty is defined as the sum in quadrature of the average and root mean square of the distribution.

Possible differences between data and simulation in the mass resolution are considered.
The mass fit to data is repeated with the parameters describing the mass resolution of the signal and of the \BsToDstD and  \BdToDstDs background contributions (widths of the Crystal Ball functions, fixed to the values determined from simulation in the nominal model) multiplied by a common, free scaling factor. The decay-time fit to data is repeated and the variation of the \CP parameters is taken as systematic uncertainty.

A possible contribution to the mass distribution due to background from \BsToDstDs decays is also considered. The same shape and parameters for the mass PDF as for the \BdToDstDs component is taken, but the mean is shifted by
the difference of the known values of the \Bs and \Bd mesons~\cite{PDG2018}.
No significant yield for this decay mode as well as no significant variation in the signal yield is found, therefore no systematic uncertainty is associated to this background.

To assess the systematic uncertainty on the decay-time efficiency description, pseudoexperiments are generated with alternative positions of the spline knots, (0.3, 1.3, 2.2, 6.3, 10.3)\ps, chosen using simulation to provide a good alternative fit. The systematic uncertainty is found to be negligible, as expected from the small correlation between the decay-time efficiency and the \CP parameters. 

The validity of using the same decay-time efficiency function for the two \Dz final state samples is tested by producing 2500 data sets with the bootstrap method and fitting each sample once with the nominal \PDF and once with different acceptances for the two final states.
Due to the limited amount of data candidates, only 32\% of the fits with separate acceptances converge with a good covariance matrix quality, however the systematic uncertainty evaluated from those fits is found to be negligible. 
The systematic uncertainty on the  \CP parameters due to variation of the decay-time resolution, assessed with pseudoexperiments, is found to negligible.

The calibration parameters of flavour tagging are constrained in the decay-time fit within their statistical and systematic uncertainties, therefore their variation is  included in the statistical uncertainty of the \CP parameters.
Two additional sources of uncertainties related to the tagging parameters, as mentioned in 
Sec.~\ref{sec:Tagging}, are considered. The first test concerns a small deviation from a perfect linear calibration of the mistag probability of the SS tagger observed in signal simulated data.
The second test accounts for  the effect of a slight dependence of the OS tagger mistag probability on the \Bz decay time. 
Pseudoexperiments are used to evaluate the variations of the \CP parameters  with respect to the unbiased case. The systematic uncertainties corresponding to the two tests are summed in quadrature and reported in Table~\ref{tab:syst_total}. 

The systematic uncertainty related to the decay-time fit input parameters ($\tau_d$, \dmd and \DGd) is determined by varying each parameter according to its uncertainty~\cite{HFLAV16}.
A test is done to check the impact on the \CP parameters of the assumption of no \CP violation in the \BdToDstDs background.  Pseudoexperiments with a charge asymmetry of 10\% included at generation for the \BdToDstDs component are studied.
No bias is found on the \CP parameters, as a consequence no systematic uncertainty is assigned due to this source.

The systematic uncertainty on \ADstD is evaluated as the weighted average of the quadratic sum of the uncertainties on the detection asymmetry, from Table~\ref{tab:asymmetries} and the uncertainties on the raw asymmetries, from Table~\ref{tab:syst_total}.

\section{Results and conclusion}\label{sec:conclusion}
A flavour-tagged decay-time-dependent analysis of \BdToDstD decays is performed using $pp$ collision data collected by the \lhcb experiment between 2011 and 2018, corresponding to an integrated luminosity of about 
9\invfb .
The \Dm meson is reconstructed as \Kp\pim\pim, while the \Dstarp meson is reconstructed as \decay{\Dstarp}{\Dz\pip}, where the $\Dz$ meson final states \Km\pip and \Km\pip\pip\pim are considered. In total, about 6,160 signal decays are selected.
Opposite-side and same-side tagging algorithms are used to determine the flavour of the \Bz mesons at production, with a total tagging power of 5.6 to 7.1\%.
The following \CP parameters are measured
\begin{eqnarray*}
    \SDstD &=     {-0.861} \pm {0.077} \,\text{(stat)} \pm {0.019} \,\text{(syst)}\,, \\
    \Delta\SDstD &=       \phantom{+}     {0.019} \pm {0.075} \,\text{(stat)} \pm {0.012} \,\text{(syst)} \,,\\ 
    \CDstD &=     {-0.059} \pm {0.092} \,\text{(stat)} \pm {0.020} \,\text{(syst)} \,,\\
    \Delta\CDstD &=           {-0.031} \pm {0.092} \,\text{(stat)} \pm {0.016} \,\text{(syst)} \,,\\ 
    \ADstD &=         \phantom{+}  {0.008} \pm {0.014} \,\text{(stat)} \pm  0.006 \,\text{(syst)}\,. 
\end{eqnarray*}
The largest statistical correlations are found between the \SDstD and \CDstD parameters and the \DeltaSDstD and \DeltaCDstD. They amount to \mbox{$\rho(\SDstD, \CDstD)=0.44$} and \mbox{$\rho(\DeltaSDstD, \DeltaCDstD)=0.46$}, respectively. 

This measurement using \BdToDstD decays excludes the hypothesis of \CP conservation at more than 10 standard deviations, obtained using Wilk's theorem~\cite{wilks1938}. 
This result is the most precise single measurement of the \CP parameters in \BdToDstD decays and it is  compatible with previous measurements by the Belle~\cite{Rohrken:2012ta} and BaBar~\cite{Aubert:2008ah} experiments. 
The precision of $\Delta\CDstD$ and \CDstD parameters is comparable with that of previous measurements, while for \SDstD, $\Delta\SDstD$ and \ADstD, this measurement improve significantly the precision of the current world average~\cite{HFLAV16}.

\section*{Acknowledgements}
%
%
\noindent We express our gratitude to our colleagues in the CERN
accelerator departments for the excellent performance of the LHC. We
thank the technical and administrative staff at the LHCb
institutes.
We acknowledge support from CERN and from the national agencies:
CAPES, CNPq, FAPERJ and FINEP (Brazil); 
MOST and NSFC (China); 
CNRS/IN2P3 (France); 
BMBF, DFG and MPG (Germany); 
INFN (Italy); 
NWO (Netherlands); 
MNiSW and NCN (Poland); 
MEN/IFA (Romania); 
MSHE (Russia); 
MinECo (Spain); 
SNSF and SER (Switzerland); 
NASU (Ukraine); 
STFC (United Kingdom); 
DOE NP and NSF (USA).
We acknowledge the computing resources that are provided by CERN, IN2P3
(France), KIT and DESY (Germany), INFN (Italy), SURF (Netherlands),
PIC (Spain), GridPP (United Kingdom), RRCKI and Yandex
LLC (Russia), CSCS (Switzerland), IFIN-HH (Romania), CBPF (Brazil),
PL-GRID (Poland) and OSC (USA).
We are indebted to the communities behind the multiple open-source
software packages on which we depend.
Individual groups or members have received support from
AvH Foundation (Germany);
EPLANET, Marie Sk\l{}odowska-Curie Actions and ERC (European Union);
ANR, Labex P2IO and OCEVU, and R\'{e}gion Auvergne-Rh\^{o}ne-Alpes (France);
Key Research Program of Frontier Sciences of CAS, CAS PIFI, and the Thousand Talents Program (China);
RFBR, RSF and Yandex LLC (Russia);
GVA, XuntaGal and GENCAT (Spain);
the Royal Society
and the Leverhulme Trust (United Kingdom).




\addcontentsline{toc}{section}{References}
\bibliographystyle{LHCb}
\bibliography{main,standard,LHCb-PAPER,LHCb-CONF,LHCb-DP,LHCb-TDR}

\newpage
\centerline
{\large\bf LHCb collaboration}
\begin
{flushleft}
\small
R.~Aaij$^{31}$,
C.~Abell{\'a}n~Beteta$^{49}$,
T.~Ackernley$^{59}$,
B.~Adeva$^{45}$,
M.~Adinolfi$^{53}$,
H.~Afsharnia$^{9}$,
C.A.~Aidala$^{79}$,
S.~Aiola$^{25}$,
Z.~Ajaltouni$^{9}$,
S.~Akar$^{64}$
% ...

{\footnotesize \it

% ...
$ ^{9}$Universit{\'e} Clermont Auvergne, CNRS/IN2P3, LPC, Clermont-Ferrand, France\\
% ...
$ ^{25}$INFN Sezione di Milano, Milano, Italy\\
% ...
$ ^{31}$Nikhef National Institute for Subatomic Physics, Amsterdam, Netherlands\\
% ...
$ ^{45}$Instituto Galego de F{\'\i}sica de Altas Enerx{\'\i}as (IGFAE), Universidade de Santiago de Compostela, Santiago de Compostela, Spain\\
% ...
$ ^{49}$Physik-Institut, Universit{\"a}t Z{\"u}rich, Z{\"u}rich, Switzerland\\
% ...
$ ^{53}$H.H. Wills Physics Laboratory, University of Bristol, Bristol, United Kingdom\\
% ...
$ ^{59}$Oliver Lodge Laboratory, University of Liverpool, Liverpool, United Kingdom\\
% ...
$ ^{64}$University of Cincinnati, Cincinnati, OH, United States\\
% ...
$ ^{67}$Syracuse University, Syracuse, NY, United States\\
% ...
$ ^{79}$University of Michigan, Ann Arbor, United States, associated to $^{67}$\\
\bigskip
% ...
}
\end{flushleft}

%
%
%
%

\end{document}